\documentclass[11pt,english,a4paper,twoside]{article}
\usepackage[cp1251]{inputenc}
\usepackage{amsmath,amsfonts,amssymb,amscd,euscript}
\usepackage[english]{babel}

\usepackage{graphicx}
\usepackage{psfrag}
\makeatletter

\def\titlerus{\thispagestyle{empty} { } \vspace{-5mm} \noindent
\raisebox{-37pt}[\headheight][0pt]{\vbox{ \hbox to \textwidth{\hfil
\scriptsize  BULLETIN OF THE UDMURTIAN UNIVERSISTY\hfil }
\vspace{2pt} \hrule \vspace{8pt} \hbox to \textwidth{\series \hfil  \issue}
\vspace{30pt} \hbox{UDC \UDK} \vspace{3pt} \hbox{MSC \MSC}
}} \vspace{ 30pt plus 6pt }}

\def\annotationandkeywordsrus{\noindent {\small \annotationrus \par } \vspace{8pt}
\noindent {\small {\it Keywords}: \keywordsrus} \vspace{3pt} \par \hbox{Received 01.11.2011} \par \vspace{10pt}}

\@addtoreset{equation}{section}
\renewcommand{\section}{\@startsection{section}{1}{0pt}{1.3ex
plus 1ex minus .1ex}{1.3ex plus .1ex}{\bf\,\S\,}}

\renewcommand{\@begintheorem}[2]{\begin{trivlist}
\item[\hspace{\labelsep}{\bf \mbox{~~~}#1\ #2.}]}
\renewcommand{\@opargbegintheorem}[3]{\begin{trivlist}
\item[\hspace{\labelsep}{\bf \mbox{~~~}#1\ #2 {\rm (#3).}}]}
\renewcommand{\@endtheorem}{\end{trivlist}}

\newtheorem{teo}{Theorem}
\newtheorem{pre}{Proposition}
\newtheorem{df}{Definition}
\newtheorem{zam}{Remark}
\newcommand{\doc}{\mbox{P r o o f}}
\renewcommand{\@evenfoot}{}
\renewcommand{\@oddfoot}{}

\renewcommand*{\@biblabel}[1]{#1.\hfill}

\newcommand*{\CSep}{.\ }
\renewcommand{\@makecaption}[2]{%
  \vskip\abovecaptionskip
  \sbox\@tempboxa{{\bf #1\CSep}{#2}}%
  \ifdim \wd\@tempboxa >\hsize
  \begin{center}%
    {\footnotesize{\bf #1\CSep}{#2\par}}%
  \end{center}%
  \else
    \global \@minipagefalse
    \hb@xt@\hsize{\hfil\box\@tempboxa\hfil}%
  \fi
  \vskip\belowcaptionskip%
}

\renewcommand{\@evenhead}{\raisebox{0pt}[\headheight][0pt]{\vbox{\hbox to\textwidth{\thepage \strut \hfil
\text{\autorsrus} \hfil } \hrule \vspace{8pt} \hbox to \textwidth{\series \hfil  \issue}}}}

\renewcommand{\@oddhead}{\raisebox{0pt}[\headheight][0pt]{\vbox{\hbox to\textwidth{  \strut \hfil
\text{\articleshortname} \hfil \thepage} \hrule \vspace{8pt} \hbox to \textwidth{\series \hfil  \issue}}}}

\newcommand{\series}{MATHEMATICS}

\newcommand{\issue}{2011. Issue\,4} 

\newcommand{\autorsrus}{M.\,P.~Kharlamov, P.\,E.~Ryabov} 

\newcommand{\articleshortname}{Smale\,--\,Fomenko diagrams}

\newcommand{\UDK}{517.938.5+531.38} 
\newcommand{\MSC}{70E17, 70G40} 

\newcommand{\annotationrus}{We present the complete analytical classification of the atoms arising at the critical points of rank~1 of the Kowalevski--Yehia gyrostat.
To classify the Smale--Fomenko diagrams,  all separating values of the gyrostatic momentum are found. We present a kind of constructor of the Fomenko graphs; its application gives the complete description of the rough topology of this integrable case. It is proved that there exists exactly nine groups of identical molecules (not considering the marks). These groups contain 22 stable types of graphs and 6 unstable ones with respect to the number of critical circles on the critical levels.}

\newcommand{\keywordsrus}{gyrostat, Kowalevski\,--\,Yehia case, diagrams,
topological invariants.}

\newcommand{\bR}{\mathbb{R}}
\newcommand{\la}{{\lambda}}
\newcommand{\mtA}{\mathbb{A}}
\newcommand{\mtB}{\mathbb{B}}
\newcommand{\mtC}{\mathbb{C}}
\newcommand{\mtD}{\mathbb{D}}
\newcommand{\mtE}{\mathbb{E}}
\newcommand{\mtF}{\mathbb{F}}
\newcommand{\mtG}{\mathbb{G}}
\newcommand{\mtH}{\mathbb{H}}
\newcommand{\ri}{\mathrm{i}\,}
\newcommand{\rk}{\mathop{\rm rank}\nolimits}
\newcommand{\sgn}{\mathop{\rm sgn}\nolimits}

\newcounter{myta}
\newcommand{\myt}{\refstepcounter{myta}\themyta}

\begin{document}

\titlerus
\begin{flushleft}
{\bf \copyright { \textit { \autorsrus}} \\[2ex]
{Smale\,--\,Fomenko diagrams and rough topological invariants\\ of the Kowalevski\,--\,Yehia case} \footnote{The work is supported by the RFBR (grants 10-01-00043, 10-01-97001).} }
\end{flushleft}

\annotationandkeywordsrus

\begin{flushleft}{\bf{Introduction}}\end{flushleft}

The paper deals with the integrable case of Kowalevski\,--\,Yehia \cite{YehEng} and continues the articles \cite{Udgu1,Udgu2}, the results of which are used below. In the work \cite{Udgu2} one can find the sufficient list of publications devoted to this problem. We study the system of equations
\begin{equation}\label{eq1_1}
\begin{array}{lll}
2\dot\omega _1   = \omega _2 (\omega _3- \la)  , &
2\dot\omega _2 =  - \omega _1 (\omega _3-\la)  - \alpha _3 , &
\dot\omega _3   = \alpha _2, \\
\dot\alpha _1   = \alpha _2 \omega _3  - \alpha _3 \omega _2 , &
\dot\alpha _2   = \alpha _3 \omega _1  - \alpha _1 \omega _3 , &
\dot\alpha_3   = \alpha_1 \omega_2  - \alpha_2 \omega_1
\end{array}
\end{equation}
on the phase space $P^5=\mathbb{R}^3({\boldsymbol\omega}){\times}S^2({\boldsymbol\alpha})$ defined in $\mathbb{R}^6$ as the level $\Gamma=1$ of the geometrical integral
$\Gamma=\alpha_1^2+\alpha_2^2+\alpha_3^2.$ The system has the following integrals in involution
\begin{equation}\label{eq1_3}
\begin{array}{l}
L = \omega _1 \alpha _1  + \omega _2 \alpha _2  +\displaystyle{\frac{1}{2}} (\omega _3+\la) \alpha _3, \qquad H = \omega _1^2  + \omega _2^2 + \displaystyle{\frac{1}{2}}\omega _3^2 -
\alpha _1, \\[2mm]
K=(\omega_1^2-\omega^2_2+\alpha_1)^2+(2\omega_1\omega_2+\alpha_2)^2 + 2\la[(\omega_3-\la)(\omega_1^2+\omega^2_2)+2\omega_1 \alpha_3].
\end{array}
\end{equation}
Therefore, the integral mapping (the momentum mapping) of the system \eqref{eq1_1} is defined as
\begin{equation}\notag
       J=L{\times}H{\times}K: P^5 \to \mathbb{R}^3.
\end{equation}
Let us denote by $\Sigma$ the bifurcation diagram of $J$. Then $\Sigma$ is a proper subset in the union of three {\rm (}intersecting{\rm )} \textit{bifurcation} surfaces $\Pi_j$ $(j=1,2,3)$
in the space $\mathbb{R}^3{(\ell,h,k)}$ of the integral constants:
\begin{eqnarray}
& \begin{array}{l}
\Pi_1 =
\left\{ \displaystyle{h= \frac{\ell^2}{s^2}+\frac{\la^2}{2} + s,}\;
\displaystyle{k=\frac{\ell^4}{s^4} - \frac{2 \ell^2}{s}+1}, \; \ell s \neq 0
\right\} \bigcup \\[3mm]
\phantom{\Pi_1 = } \quad \bigcup \left\{k=1, \; \ell=0 \right\} \bigcup \left\{\displaystyle{k=1+(h-\frac{\la^2}{2})^2},\;  \ell=0 \right\},
\end{array} \nonumber\\
& \Pi_{2,3} = \left\{ \displaystyle{h=2 \ell^2 + \frac{1}{2 s} - \frac{\la^2}{2}(1-4s^2),}\;
k=-4 \ell^2 \la^2 + \displaystyle{\frac{1}{4 s^2}} - \displaystyle{\frac{\la^2}{s}(1-\la^2s)(1-4s^2)}
\right\}.\nonumber
\end{eqnarray}
Here $s <0 $ for $\Pi_2$ and $s > 0 $ for $\Pi_3$.
The complete investigation of the conditions defining $\Sigma$ in this union for all values of the parameters is given in \cite{mtt40, Udgu2}.

The function $L$ is a Casimir function for the Poisson brackets on $P^5.$ Therefore, on each level
${P^4_\ell} = \{L=\ell\} \subset P^5$ the induced vector field is a Hamiltonian system with two degrees of freedom. The integral map
${\mathcal{R}_\ell=H{\times}K|_{{P^4_\ell}}:{P^4_\ell} \to \mathbb{R}^2}$ makes this system completely integrable, so all notions and results of the general theory \cite{FomDAN86,BolFom} are applicable. Obviously, the bifurcation diagram of the map $\mathcal{R}_\ell$ is the cross-section of the set $\Sigma$ by the plane $\ell={\rm const}$.

The set $\mathcal{C}$ of critical points of the momentum map $J$ is stratified by the rank of $J$. Since the integral $L$ is everywhere regular and foliates $P^5$ into smooth symplectic leaves ${P^4_\ell}$, it is natural to accept the following terminology.
\begin{df}\label{def1}
Let $x \in \mathcal{C} \subset P^5$. Then it belongs to ${P^4_\ell}$ for a certain $\ell$. We call the rank and the type of $x$ with respect to the induced map $\mathcal{R}_\ell$ the rank and the type of this point in $P^5$. A critical point is said to be degenerate or non-degenerate if it is, respectively, degenerate or non-degenerate in the corresponding subsystem on ${P^4_\ell}$.
\end{df}
Thus, the rank of a critical point $x$ is by definition equal to $\rk J(x)-1$.
The bifurcation diagram $\Sigma$ is a $J$-image of the set of critical points of rank~0 and 1.

As in any system with symmetry, iso-energetic manifolds are supplied with two indices \mbox{$\ell$ and $h$}:
$${Q_{\ell,h}^3} =\{x\in P^5: L(x)=\ell, H(x)=h\}.
$$
In our case these manifolds depend also on the parameter $\la$: ${Q_{\ell,h}^3}={Q_{\ell,h}^3}(\la).$ We call an iso-energetic manifold ${Q_{\ell,h}^3}$ \textit{typical} if it contains no critical points of rank~0 and no degenerate critical points of rank~1. A point $(\ell,h)$ corresponding to a typical ${Q_{\ell,h}^3}$ is called a {\it typical} point of the plane $O\ell h$. The topology of the Liouville foliation arising on a typical ${Q_{\ell,h}^3}$ up to the rough equivalence is described by the corresponding Fomenko invariant \cite{FomFAN88}; this invariant is also called the Fomenko graph or the molecule \cite{BolFom}. The aim of this paper is to give the complete classification of such molecules as a basis for the future {\it precise} topological analysis with the help of \textit{marked} invariants \cite{FomZ}. The majority of molecules in this problem was obtained earlier in the works \cite{RyabDis,RyabRCD}. In the work \cite{GashMttTop} the classification of the Fomenko graphs was presented without the condition of \textit{identity} of the corresponding molecules in the sense of \cite{BolFom}.

\begin{flushleft}
{\bf{\S\,1. Smale's diagrams and iso-energetic manifolds }}
\end{flushleft}

A gyrostat is a mechanical system with four degrees of freedom (a rigid body with a fixed point and a symmetric rotor with the axis of rotation fixed in the body). For this system, $\la$ is the constant of a cyclic integral and is naturally included in the set of integral constants. Therefore, all statements about the properties preserved in the space of the integral parameters naturally hold with respect to the enhanced space of parameters including the axis $\mathbb{R}=\mathbb{R}_\la$. According to this, we agree on the following terminology. Let $A$ be some set and $B(\la)$ be a family of its subsets depending on the parameter $\la$. We then put $\widehat{A} = A {\times} \mathbb{R}_\la$ and denote by $\widehat{B}$ the union of the subsets $B(\la)$ in the $\la$-sections of~$\widehat{A}$. If the term ``object'' is assigned to a set $A$ or $B(\la)$, then the corresponding $\widehat{A}$ or $\widehat{B}$ will be called the ``enhanced object''.

In equations \eqref{eq1_1}, we are still free to choose the directions of the movable axes. In the sequel, we suppose these directions to be chosen in such a way that ${\la \geqslant 0}$. The case ${\la=0}$ corresponds to the classical Kowalevski problem completely investigated previously in \cite{KhPMM83,KhDan83}. This case is considered here only as a limit case to compare the obtained results.

The topological type of ${Q_{\ell,h}^3}$ changes at the points of the bifurcation diagram ${\Sigma_{LH}=\Sigma_{LH}(\la)}$ of the map $L{\times}H$. The set $\Sigma_{LH}$ is called Smale's diagram. In the {\it enhanced} space $\mathbb{R}^3{(\ell,h,\la)}$ we obtain {\it enhanced} Smale's diagram $\widehat{{\Sigma_{LH}}}$. The latter divides $\mathbb{R}^3{(\ell,h,\la)}$ into open connected components, which are usually called \textit{chambers}. The symmetry of the phase space of this problem $(\omega_1,\omega_2,\alpha_3) \mapsto (-\omega_1,-\omega_2,-\alpha_3)$ establishes an isomorphism of the flows on ${Q_{\ell,h}^3}$ and ${Q_{-\ell,h}^3}$. Due to this fact we consider the union of two components differing by the sign of $\ell$ as one chamber.

The equations of Smale's diagrams are given in \cite{RyabRCD} (see also \cite{RyabDis}), the topological types of ${Q_{\ell,h}^3}$ are found in \cite{RyabRCD,GashMttTop}. All types of Smale's diagrams and the existing chambers are defined in \cite{RyabDis} with the help of numerical methods. In Theorem~\ref{th1} below we give the strict analytical basis for this classification.

\begin{zam}\label{rem1}
Let us agree on one more system of notation. Let $\Phi$ be a certain subset of the phase space. Mostly, it will be a set of critical points of a given type. On this set the general integrals \eqref{eq1_3} are defined, but also some partial integrals can exist. Then we obtain some map of the set $\Phi$ into the space or the plane of some integral constants. If it is clear from the context what integral map is considered, then we use the same notation for the image of the set $\Phi$, thus obtaining the surface or the curve denoted by $\Phi$. For a given set $S$ in the space or the plane of integral constants we call a point $s\in S$ \textit{admissible} if the inverse image of $s$ contains real solutions of equations \eqref{eq1_1} of the considered type, e.g., motions belonging to the considered critical subsystem or any real solutions if the whole system \eqref{eq1_1} is meant.
\end{zam}

\begin{teo}\label{th1}
{\it In the Kowalevski\,--\,Yehia problem there exist seven types of Smale's diagrams ${\Sigma_{LH}}(\la)$ stable with respect to the parameter $\la$. The separating values of the parameter are}
$0,$ $\la_1,$ ${\la_*=2^{-3/4}},$ ${\la^*=(4/3)^{3/4}},$ $2\sqrt{\sqrt{2}-1}$, $\la_2,$ $\sqrt{2}$, {\it where}
\begin{equation}\notag
\begin{array}{lll}
  \la_1=\displaystyle{\frac{(X-4)^{3/2}}{2 \sqrt{2} X^{3/4}}} \approx 0.023, & X^4 - 24 X^3 + 720 X^2 - 2048 X - 3072 = 0, & X \approx 4.342,  \\[3mm]
  \la_2=\displaystyle{\frac{3X-4}{2 X^{3/4}}} \approx 1.326, &  3 X^4 + 32 X^3 - 180 X^2 + 96 X - 64 = 0, & X \approx 3.685.
\end{array}
\end{equation}
{\it The enhanced diagram $\widehat{{\Sigma_{LH}}}$ divides the space $\mathbb{R}^3{(\ell,h,\la)}$ into eight chambers
$\mtA,\ldots, \mtH$ with nonempty iso-energetic manifolds. The existence conditions for the chambers with respect to $\lambda,$ the number of connected components in the chambers and the topology of ${Q_{\ell,h}^3}$ are presented in Table~{\rm \ref{table1}}. All iso-energetic manifolds are connected sets, notation $K^3$ stands for the connected sum $(S^2{\times}S^1)\#(S^2{\times}S^1)$.}\end{teo}

\def\rul{\rule[-3pt]{0pt}{14pt}}

\begin{table}[ht]
\centering
\small

\begin{tabular}{|c| l| c| c|}
\multicolumn{4}{r}{{Table \myt\label{table1}}}\\
\hline
\begin{tabular}{c}{Chamber}\\[-3pt]{code}\end{tabular} &\begin{tabular}{c}{Life time}\\[-3pt]{w.r.t. $\la$}\end{tabular}
&\begin{tabular}{c}{Components}\\[-3pt]{in the chamber}\end{tabular}&\begin{tabular}{c}${Q_{\ell,h}^3}$\end{tabular} \\
\hline
$\rul\mtA$ & $\la\in[0,+\infty)$ & 1 & $S^3$\\
\hline
$\rul\mtB$ & $\la\in[0,+\infty)$ & 2 & $S^2{\times}S^1$\\
\hline
$\rul\mtC$ & $\la\in(0,\la_*)$ & 2 & $S^2{\times}S^1$\\
\hline
$\rul\mtD$ & $\la\in[0,\la_1)$ & 2 & $K^3$\\
\hline
$\rul\mtE$ & $\la\in[0,+\infty)$ & 1 & ${\mathbb{R} P^3}$\\
\hline
$\rul\mtF$ & $\la\in(\la_*,\la_2)$ & 2 & $S^2{\times}S^1$\\
\hline
$\rul\mtG$ & $\la\in(\la^*,\sqrt{2})$ & 1 & $K^3$\\
\hline
$\rul\mtH$ & $\la\in(\la^*,+\infty)$ & 2 & $S^2{\times}S^1$\\
\hline
\end{tabular}
\normalsize
\end{table}

To prove the theorem it is sufficient to point out that Smale's diagram is the image of the singular points of the Euler--Poisson equations; the set of these points $\mathcal{C}^0$ can be parametrized by the axial component of the angular velocity $\omega_3=r$~\cite{GashMttTop}. The values of the first integrals on $\mathcal{C}^0$ (the equations of the diagram) are as follows
\begin{equation}\label{eq2_7}
\begin{array}{l}
    \ell = \mp \displaystyle{\frac{1}{2}[\la(r-\la)+d]}\sqrt{\displaystyle{\frac{r}{2}\left[-r+\frac{1}{r-\la}d\right]}},\quad
    h= -\displaystyle{\frac{1}{2}r(r-\la)+\frac{2r-\la}{2(r-\la)}d}, \\
    d^2 = 4+r^2(r-\la)^2, \qquad  r \in (-\infty,0] \cup [0,\la) \cup (\la,+\infty).
\end{array}
\end{equation}
The sign of $d$ for $r\ne 0$ is defined as $\mathop{\rm sgn}\nolimits d =\mathop{\rm sgn}\nolimits \bigl[r (r-\la) \bigr]$ and is arbitrary if $r=0$. In particular, for any $\la$ the value $r=0$ corresponds to two points of the phase space; these points are the absolute equilibria of the body ${\boldsymbol\omega}=0,\  {\boldsymbol\alpha}=(\pm 1,0,0)$. This fact explains the choice of the segments for $r$ (zero is included twice). We denote the subsets in $\mathcal{C}^0$ by $\delta_1$ for $r\in [0,\la)$, $\delta_2$ for $r\in (-\infty,0]$,  $\delta_3$ for $r\in (\la, +\infty)$ (the latter set consists of two connected components). According to Remark~\ref{rem1}, the same symbols $\delta_j$ stand for the images of these sets, i.e., \textit{the curves} $\delta_j$ in the spaces of the integral constants or {\it the surfaces} $\delta_j$ in the enhanced spaces (these surfaces are then generated by the parameters $r,\la$).

\begin{figure}[ht]
\centering
\includegraphics[width=100mm,keepaspectratio]{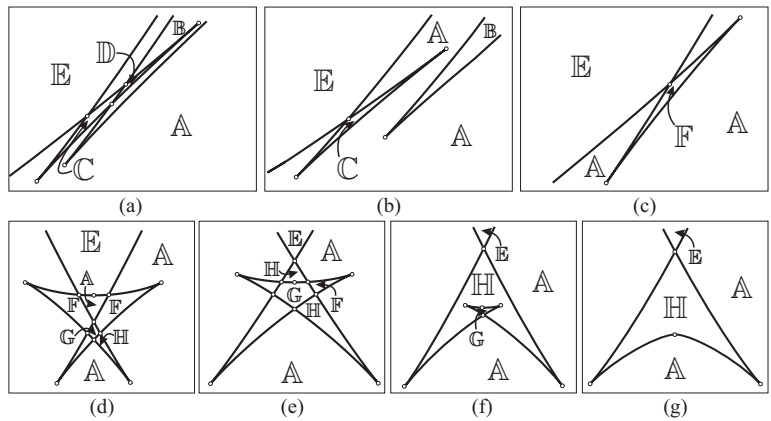}\\
\caption{The fragments of Smale's diagrams and the chambers}\label{fig_smalechambs}
\end{figure}

The evolution of the curves $\delta_j$ with respect to $\la$ can be easily investigated analytically. On Fig.~\ref{fig_smalechambs} for the intervals between the separating values of $\la,$ we show the fragments of the diagrams in which the changes take place. Also on Fig.~\ref{fig_smalechambs} the notation of the arising chambers is presented. The curve $\delta_1$ does not take part in the changes and remains the lower boundary for the admissible values of $h$. When we cross the value $\la_*$, i.e., pass from $(b)$ to $(c)$, the cusps on the curve $\delta_2$ first glue together with the self-intersection point and then part again, the chamber $\mtC$ disappears, the new chamber $\mtF$ is born. When $\la>\la_*$ the curve $\delta_3$  does not take part in the changes any more. When $\la$ crosses the values $\la^*,2\sqrt{\sqrt{2}-1},\sqrt{2}$ the changes take place on the axis $\ell=0$. The separating values $\la_1,\la_2$ are found from the condition that a cusp of the curve $\delta_2$ belongs to the curve $\delta_3$ or a cusp of $\delta_2$ coincides with another, regular, point of $\delta_2$. In the statement of the theorem we show the substitutions of $\la_{1,2}$ in terms of some new variable $X$ leading to the equations in $X$, each equation having exactly one real root in the interval needed to guarantee the positive value of $\la$.

The topology of ${Q_{\ell,h}^3}(\la)$ is defined according to S.\,Smale as a reduced tangent bundle (reduced bundle of circles in tangent planes) of the corresponding region of possible motions on the Poisson sphere $\{{\boldsymbol\alpha}: U_{\ell,\la}({\boldsymbol\alpha}) \leqslant h\}$ and is found with the methods of the Morse theory. Here $U_{\ell,\la}$ is the amended potential. Note that for any function $f$ on $\mathbb{R}^3$ the characteristic polynomial of the second differential of the restriction of $f$ to the unit sphere is
\begin{equation}\notag
\xi_f(\mu) = \frac{1}{\mu}\det \bigl[\Theta^2 f - \mu E\bigr],\qquad \Theta = {\boldsymbol\alpha}\times \frac{\partial}{\partial {\boldsymbol\alpha}}.
\end{equation}
For $f=U_{\ell,\la}$ the roots of this polynomial (the Morse characteristic values) at the points of $\mathcal{C}^0$ are as follows:
\begin{equation}\notag
    \begin{array}{l}
    \mu_1 = -\displaystyle{\frac{1}{2}\left[ r(r-\la)+d\right]}, \quad \mu_2=-\displaystyle{\frac{1}{2(r-\la)d}\left[ (2r-\la)(r-\la)- d\right]\left[ (2r-\la)(r-\la)r + \la d\right]}.
    \end{array}
\end{equation}
In particular, $\mathop{\rm sgn}\nolimits \mu_1 = - \mathop{\rm sgn}\nolimits d$. Therefore, $\mu_1>0$ on $\delta_1$ and $\mu_1<0$ on $\delta_2, \delta_3$. The value $\mu_2$ is positive on the curve $\delta_1.$ For all other cases, the sign of $\mu_2$ is defined according to the agreement about the sign of $d$ from the position of the point $(r,\la)$ in the domain $r \ne \la$ with respect to the curves $\mu_2(r,\la)=0$. These curves, obviously, correspond to the cusps on Smale's diagrams.

\begin{flushleft}
{\bf{\S\,2. Critical subsystems}}
\end{flushleft}

Recall the notion of a critical subsystem \cite{KhRCD05,KhND07}. In the problem considered we say that the critical subsystem ${\mathcal{M}}_j$ is the set of critical points of the momentum map belonging to the inverse image of the bifurcation surface $\Pi_j$ $(j=1,2,3).$ In the neighborhood of saddle type critical points of rank~0 we have to be more accurate. Let us write the equation of $\Pi_j$ in the form
$$
P_j(\ell,h,k)=0,
$$
where $P_j$ is an irreducible polynomial. Substituting $\ell,h,k$ with the general integrals $L,H,K$ we obtain the function $f_j=P_j(L,H,K)$ on the phase space. Then we define the critical subsystem ${\mathcal{M}}_j$ as the set of critical points of $f_j$ belonging to the zero level $f_j=0$.

Let $(\ell,h)$ be a {\it typical} point. Then the intersection ${Q_{\ell,h}^3}\cap \mathcal{C}$ consists of a finite number of non-degenerate critical circles. Consider the straight line $\tau_{\ell,h} \subset \mathbb{R}^3{(\ell,h,k)}$ over $(\ell,h)$ parallel to $Ok$. In this case critical circles correspond to a finite number of {\it transversal} intersections of $\tau_{\ell,h}$ with the surfaces $\Pi_j$. To build the molecule, we need to find the atoms corresponding to these intersections and, for non-symmetric atoms, establish their orientation with respect to the direction of growth of the integral $K$. To answer practically all questions, one has to take a critical point $x$ in the inverse image of the intersection and calculate its type and the Morse\,--\,Bott index, i.e., the index of the second differential of the function $K$ restricted to the \textit{transversal} subspace $T_{\ell,h}(x)$ drawn inside ${Q_{\ell,h}^3}$ at the point $x$ to the critical circle containing $x$. It is known that transversal intersections of the surfaces $\Pi_j$ correspond to non-degenerate critical points of rank~0, while tangency lines of the surfaces have degenerate critical points of rank~1 in the inverse images \cite{Udgu1}. Hence the image of any critical point $x$ of the function $K$ on a typical ${Q_{\ell,h}^3}$ can belong only to one surface, and the point $x$ itself can belong only to one critical subsystem.
Thus, we have to classify the points of the critical subsystems using as a separating set the set of non-generic critical points. For each of the arising domains we then have to calculate the points type and the Morse\,--\,Bott index. In a cross-section of the constant value of $\ell$ or $h$ of the bifurcation diagram $\Sigma \subset \mathbb{R}^3{(\ell,h,k)}$ each of the obtained domains gives an arc, along which the corresponding atom of the molecule $W_{\ell,h}$ has the same type. In the enhanced space on the surface of a cross-section of the three-dimensional complex $\widehat{\Sigma}$ by the corresponding hyperplane we obtain domains, in which the type and orientation of the atom included in the molecule $W_{\ell,h}(\la)$ are preserved. Let us turn to the classification of points in the critical subsystems.

The first critical subsystem is defined by the equations \cite{PVLect}
\begin{equation}\notag
{\mathcal{M}_1}: \left\{    \begin{array}{l}
      \omega_1 = p , \qquad  \omega_2 = 0, \qquad \omega_3 = r,\\
      \displaystyle{\alpha_1=\frac{1}{2} r^2+p ^2 - h, }   \qquad      \displaystyle{\alpha_2=\sqrt{R(r)}, } \qquad \alpha_3 = - p  (r - \la ).
    \end{array} \right.
\end{equation}
Here
$$
\displaystyle{p^2=h-\frac{\la^2}{2}-s}, \quad \ell=- s\,p,\quad \displaystyle{R=-\frac{1}{4}r^4-(2p ^2-h)r^2+2 \la p ^2 r+1-(p ^2-h)^2-p ^2\la^2}
$$
and ${\dot r}=\sqrt{R(r)}.$ For the coordinates on ${\mathcal{M}_1}$ one may choose $r$ (the variable along critical circles) and two of the integral constants $s,h,\ell$. Suppose that for given $s,\ell,h$ satisfying the equations of $\Pi_1$ the polynomial $R(r)$ has no multiple roots. It means that there are no critical points of rank~0 on a given integral level. Then, obviously, the number of periodic solutions on this level in $\mathcal{M}_1$ is equal to the number of intervals on which $R(r)$ is positive.

The type of a critical point of rank~1 of the subsystem ${\mathcal{M}_1}$ is defined by the symplectic operator generated by the function $F_1=K-2p^2 H$ \cite{Udgu1}. The eigenvalues of this operator are $\pm \sqrt{m_1}$, where
\begin{equation}\notag
    \begin{array}{l}
       m_1=2\displaystyle{\left[2 s^2 - 2 (h + \frac{\la^2}{2}) s+1 \right]}\displaystyle{\left[\frac{3}{2}s- (h - \frac{\la^2}{2})\right] }= \displaystyle{\frac{1}{s^3}(2\la^2 s^2-s +2\ell^2)(s^3-2\ell^2).}
     \end{array}
\end{equation}
According to the factors in this expression, let us denote the sets of degenerate critical points of rank~1 in the subsystem ${\mathcal{M}_1}$ by ${\Delta_0}$ and ${\Delta_1}$. Recalling Remark~\ref{rem1}, we use the same notation for the images of these sets in any space of integral constants. In particular, taking into account the existence conditions for critical motions found in \cite{mtt40,Udgu2}, for the integral constants $s,\ell,h$ we get
\begin{eqnarray}
      {\Delta_0} & : &\ell =\displaystyle{\pm \sqrt{\frac{s}{2}(1-2\la^2 s)}}, \quad \displaystyle{h=s+\frac{1}{2s}-\frac{\la^2}{2}}, \quad  \displaystyle{s\in (0,\frac{1}{2\la^2}]}; \label{eq3_5}\\
      {\Delta_1} & : &\ell =\displaystyle{\pm \frac{1}{\sqrt{2}}s^{3/2}}, \quad \displaystyle{h=\frac{3}{2}s +\frac{\la^2}{2}}, \quad s\in [0,s_*], \label{eq3_6}
\end{eqnarray}
where $s_*(\la)\in \mathbb{R}$ is the largest real root of the polynomial $9s^4+2\la^2 s^3-24 s^2-24 \la^2 s+4(4-\la^4)$ (positive real roots exist for all $\la$).
The first curve is the tangent line of the surfaces $\Pi_1,\Pi_3$, and the second one is the part of the cuspidal edge of the surface $\Pi_1$ between its points of intersection with the components of the curve $\delta_3$.

Note that any point in ${\mathcal{M}_1}$, if considered as a point in $\mathbb{R}^6({\boldsymbol\omega},{\boldsymbol\alpha})$, is a critical point of the function
$K_1=K-2p^2 H - 4p L-\Gamma$ (calculating the differential $dK_1$ we suppose $p$ to be constant). Hence, to find two characteristic Morse\,--\,Bott values one needs to write out the characteristic polynomial of the restriction of $d^2K_1$ to $T_{\ell,h}$. On each trajectory there is a point $x_0$ at which $R(r)=0$. At this point we can take the following basis in  $T_{\ell,h}(x_0)$:
\begin{equation*}
  v_1 = \bigl(0,1,0,0,0,0\bigr), \qquad v_2=\displaystyle{\bigl(\la+r,0,-4p ,2p (\la-r),0,2(h- p ^2-\frac{r^2}{2})\bigr)}.
\end{equation*}
The eigenvalues of the restriction of $d^2K_1$ to the span of $v_1,v_2$ are
\begin{equation*}
\mu_1 = \displaystyle{2 \left[ 2s -(\la-r)^2 \right]}, \qquad \mu_2 = \displaystyle{- 32\left(h -\frac{3}{2} s-\frac{\la^2}{2} \right)\left[h- \frac{3}{2} s -\frac{\la^2}{2} -(\la+r)^2\right]}.
\end{equation*}
In particular, by virtue of the equality $R(r)=0$, the product
\begin{equation}\notag
  \mu_1 \mu_2 = \displaystyle{-64 \left[\frac{3}{2}s- (h - \frac{\la^2}{2})\right] \left[2 s^2 - 2 (h + \frac{\la^2}{2}) s+1 \right]}
\end{equation}
does not depend on $r$ and its sign is defined by the position of the point $(s,h)$ with respect to the set ${m_1=0}$. Therefore the values $\mu_1,\mu_2$ never vanish on non-degenerate trajectories and, consequently, have constant sign.

Recall that the atom $B$ in a three-dimensional iso-energetic manifold is a direct product of a circle and a standard bifurcation of one circle into two circles through the eight line curve. According to this, the atom $B$ is essentially non-symmetric. Consider a cross-section transversal to the circle of the first multiple. The levels of $K$ surrounding the eight line curve as a whole in this cross-section (i.e., diffeomorphic to a circle that has the whole eight line curve as a limit as the value of the additional integral tends to the critical one) and the corresponding edge of the graph will be called the ``outer'' levels and the ``outer'' edge or the ``head'' of the graph $B$. The levels giving a pair of circles inside the loops of the eight line curve (i.e., each such circle tends only to one loop of the eight line curve as the value of the additional integral tends to the critical one) and the two edges of the graph corresponding to such levels will be called the ``inner'' ones or the ``legs'' of the graph $B$.

For the sake of brevity, considering the sequence of bifurcations in the direction of increasing the integral $K$ we denote the atom $A$ with the edge going up and the atom $B$ with its ``head'' down respectively by $A_+$ and $B_+$ (the number of tori increases). The atom $A$ with the edge down and the atom $B$ with its ``head'' up will be denoted respectively by $A_-$ and $B_-$ (the number of tori decreases).

\begin{pre}\label{propos2}
{\it As the integral $K$ increases on an iso-energetic level ${Q_{\ell,h}^3}$, we obtain the following bifurcations at the points of the critical subsystem ${\mathcal{M}_1}$ belonging to non-degenerate critical circles{\rm :}

{\rm 1)} for elliptic trajectories we have the atom $A_+$ if $\mu_1>0,\mu_2>0$ and the atom $A_-$ if ${\mu_1 < 0}$, $\mu_2<0;$

{\rm 2)} for one hyperbolic trajectory on a critical level of $K$ we have the atom $B_-$ if ${\mu_1 > 0}$, $\mu_2<0$ and the atom $B_+$ if $\mu_1 < 0,$ $\mu_2 > 0;$

{\rm 3)} for two hyperbolic trajectories on a critical level of $K$ with the same pairs $(\sgn \mu_1,\sgn \mu_2)$ we have two atoms $B_-$ if $\mu_1 > 0,$ $\mu_2<0$ and two atoms $B_+$ if $\mu_1 < 0,$ $\mu_2 > 0;$

{\rm 4)} for two hyperbolic trajectories on a critical level of $K$ with the opposite pairs $(\sgn \mu_1,\sgn \mu_2)$ we have two atoms $A^*.$}
\end{pre}

\doc. For elliptic trajectories the statement is obvious. It can be shown that, for hyperbolic trajectories, the vector $v_1$ is directed to the outer part of the eight line curve, because the direction of the axis $O\omega_2$ is in charge of the transfer from a critical surface to an enveloping torus. To see this, we can analyze, for example, the projections of integral manifolds onto the plane $O\omega_1 \omega_2$ (the results of such analysis are briefly presented in \cite{GashMttTop}). One can see that, similar to the classical problem,  the projection never breaks to parts in the direction of the axis $O\omega_2$. Therefore, if $\mu_1>0$, the integral $K$ on the transversal section to the critical circle increases to the outer circle and decreases to the pair of inner ones.
If on two hyperbolic critical circles the pairs $(\sgn \mu_1,\sgn \mu_2)$ are different, then supposing the existence of two atoms $B$ with the opposite directions of ``heads'' we obtain the bifurcation of three tori into three ones. As it is stated in \cite{GashMttTop}, the number of tori on a regular level can be only 1, 2 or 4. Then for the case in question the only possible bifurcation is four-to-four. However, in this problem such adjacent chambers are not found. If we suppose that we have here the atom $C_2$, then the analytical solution \cite{Gash1} must describe a heteroclinic trajectory, which also is not found. Therefore, it is the case of two atoms $A^*.$ \hfill $\square$

We see that the set of critical points of rank~0 and degenerate critical points of rank~1 serves as a separating set for the classification of atoms in critical subsystems. Let us call this set the \textit{key} set of a critical subsystem.
\begin{df}\label{def2}
Let $f$ and $g$ be two integrals of a critical subsystem independent almost everywhere. The image of the key set of this subsystem under the map $f{\times}g$ is called the $(f,g)$-diagram of the subsystem.
\end{df}

\begin{figure}[ht]
\centering
\includegraphics[width=160mm,keepaspectratio]{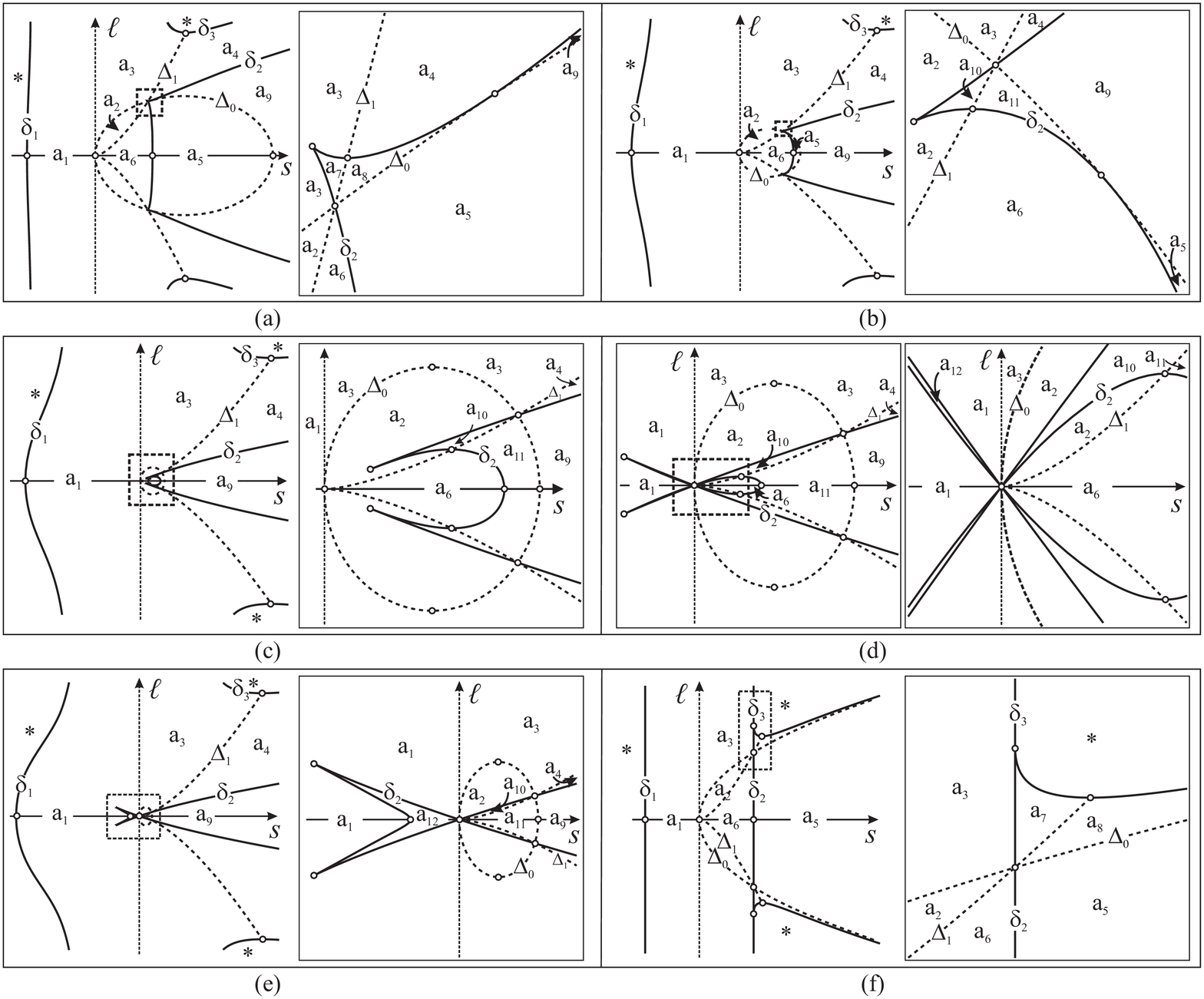}\\
\caption{The diagram of the subsystem ${\mathcal{M}_1}$}\label{fig_sys1}
\end{figure}

The $(S,L)$-diagram of the critical subsystem ${\mathcal{M}_1}$ is obtained from the bifurcation diagram consisting of the curves $\delta_1, \delta_2, \delta_3$ \cite{Udgu2} by adding the curves $\Delta_0$ and $\Delta_1$, which are the projections onto the $(s,\ell)$-plane of the curves \eqref{eq3_5}, \eqref{eq3_6}.
The transformations of the diagram for $\la>0$ take place at the following values of the parameter: $\la_*$, $1$, $\la^*$, $\sqrt{2}$. The study of intersections of the curves $\delta_2,{\Delta_0}$ gives, in addition to the values found in \cite{Udgu2}, the separating value $\la=1$. All types of the diagram of the subsystem ${\mathcal{M}_1}$ are shown in Fig.~\ref{fig_sys1}: $(a)$~$0<\la<\la_*$; $(b)$~$\la_*<\la<1$; $(c)$~$1<\la<\la^*$; $(d)$~$\la^*<\la<\sqrt{2}$; $(e)$~$\la>\sqrt{2}$. The last case $(f)$ shows, for comparison, the diagram of the classical problem $\la=0.$ In all cases except $(d)$ we show the diagram itself and its enlarged fragment. In the case $(d)$ we show, in two scales, only the fragment containing all elements that have changed with respect to the previous value $\la$. The asterisk denotes the domains that have no critical points in the inverse image of the surface $\Pi_1$. The admissible region (i.e., the region of the $(s,\ell)$-plane with nonempty critical integral manifolds) is divided into the open connected domains $a_1 - a_{12}$. Due to the mentioned above symmetry $\ell \to -\ell$, the domains symmetric to each other with respect to the $s$-axis have the same notation. In Table~\ref{table2}, we collect the  information needed to define all atoms included in Fomenko graphs for intersections with the surface~$\Pi_1,$ namely, $\la$-segment for which the domains exist in the enhanced space (life time), the number of critical circles for a domain point, the existence of common points of the domains in the enhanced space with the previously investigated zones $\la=0$ \cite{KhPMM83,KhBook88,BRF} and $\ell=0$ \cite{RyabDis,RyabRCD,Mor}. If the domain contains such common points, the last column of Table~\ref{table2} gives the notation of the corresponding arcs, paths or graphs in the cited papers containing such an atom. We see that the only domain without an analogue is $a_4$; the critical circles in $a_4$ are of elliptic type and the Morse\,--\,Bott index equals zero. Therefore, in fact, all information here follows from the previous investigations.

\begin{table}[ht]
\centering
\small

\begin{tabular}{|c|c| c| c| c|c|}
\multicolumn{6}{r}{{Table \myt\label{table2}}}\\
\hline
\begin{tabular}{c}{Domain}\\[-3pt]{(life time)} \end{tabular}
&\begin{tabular}{c}{Critical}\\[-3pt]{circles}\end{tabular}&\begin{tabular}{c}{Morse\,--\,Bott}\\[-3pt]{char. vals.}\end{tabular}
&\begin{tabular}{c}{Exit to }\\[-3pt]{$\la=0$/$\ell=0$}\end{tabular}
&\begin{tabular}{c}{Atom}\end{tabular} &\begin{tabular}{c}{Analogues}\end{tabular}\\

\hline
\begin{tabular}{c}$a_1$\\($ 0 \leqslant \la <+\infty$) \end{tabular} &{1}& {($-\;-$)} & Yes/Yes &$A_-$ & \begin{tabular}{l} $2,3$ \cite[Fig.\,6.3]{KhBook88}\\ $a_1, a_2$ \cite[Fig.\,2]{RyabRCD} \\$\gamma_1,\gamma_4$ \cite[Fig.\,11]{BRF}\\$\alpha_2,\alpha_3$ \cite[Fig.\,1]{Mor}\end{tabular}\\

\hline
\begin{tabular}{c}$a_2$\\($ 0 \leqslant \la <+\infty$) \end{tabular} &{1}& {($-\;-$)} & Yes/Yes &$A_-$ & \begin{tabular}{l} $3,3^\prime$ \cite[Fig.\,6.3]{KhBook88}\\ $a_2$ \cite[Fig.\,2]{RyabRCD} \\$\gamma_1,\gamma_4$ \cite[Fig.\,11]{BRF}\\$\alpha_2,\alpha_3$ \cite[Fig.\,1]{Mor}\end{tabular}\\

\hline
\begin{tabular}{c}$a_3$\\($ 0 \leqslant \la <+\infty$) \end{tabular} &{1}& {($+\;-$)} & Yes/No &$B_-$ & \begin{tabular}{l} 9 \cite[Fig.\,6.3]{KhBook88}\\$\gamma_5$ \cite[Fig.\,11]{BRF}\end{tabular}\\

\hline
\begin{tabular}{c}$a_4$\\($ 0 < \la < +\infty$) \end{tabular} &{1}& {($+\;+$)} & No/No &$A_+$ & {}\\

\hline
\begin{tabular}{c}$a_5$\\($ 0 \leqslant \la < 1 $) \end{tabular} &{2}& {\begin{tabular}{c}($+\;-$),($-\;+$)\end{tabular}} & Yes/Yes &$2A^*$ & \begin{tabular}{l} 6 \cite[Fig.\,6.3]{KhBook88}\\ $a_4$ \cite[Fig.\,2]{RyabRCD}\\$\gamma_2$ \cite[Fig.\,11]{BRF}\\$\delta_1,\delta_2$ \cite[Fig.\,1]{Mor} \end{tabular}\\

\hline
\begin{tabular}{c}$a_6$\\($ 0 \leqslant \la <\sqrt{2} $) \end{tabular} &{1}& {($-\;+$)} & Yes/Yes &$B_+$& \begin{tabular}{l} 5  \cite[Fig.\,6.3]{KhBook88}\\ $b_2$ \cite[Fig.\,3]{RyabRCD}\\$\gamma_3$ \cite[Fig.\,11]{BRF}\\$\beta_1$ \cite[Fig.\,1]{Mor} \end{tabular}\\

\hline
\begin{tabular}{c}$a_7$\\($ 0 \leqslant \la < \la_*$) \end{tabular} &{2}& {\begin{tabular}{c}($+\;-$),($+\;-$)\end{tabular}} & Yes/No &$2B_-$& \begin{tabular}{l} D\,\cite[Fig.\,2]{KhDan83}\\$\gamma_6$ \cite[Fig.\,11]{BRF}\end{tabular}\\

\hline
\begin{tabular}{c}$a_8$\\($ 0 \leqslant \la < \la_*$) \end{tabular} &{2}& {\begin{tabular}{c}($+\;+$),($+\;+$)\end{tabular}} & Yes/No &$2A_+$& \begin{tabular}{l} E\,\cite[Fig.\,2]{KhDan83}\\$\gamma_7$ \cite[Fig.\,11]{BRF}\end{tabular}\\

\hline
\begin{tabular}{c}$a_9$\\($ 0 < \la <+\infty$) \end{tabular} &{2}& {\begin{tabular}{c}($+\;+$),($-\;-$)\end{tabular}} & No/Yes &$A_+, A_-$& \begin{tabular}{l} $a_5$ \cite[Fig.\,2]{RyabRCD}\\$\alpha_5,\alpha_6$ \cite[Fig.\,1]{Mor} \end{tabular}\\

\hline
\begin{tabular}{c}$a_{10}$\\($ \la_* < \la <+\infty$) \end{tabular} &{2}& {\begin{tabular}{c}($-\;-$),($-\;-$)\end{tabular}} & No/Yes &$2A_-$& \begin{tabular}{l} $c_3, c_4$ \cite[Fig.\,4]{RyabRCD}\\$\alpha_3,\alpha_8$\cite[Fig.\,1]{Mor} \end{tabular}\\

\hline
\begin{tabular}{c}$a_{11}$\\($ \la_*< \la < +\infty $) \end{tabular} &{2}& {\begin{tabular}{c}($-\;+$),($-\;+$)\end{tabular}} & No/Yes &$2B_+$& \begin{tabular}{l} $b_4$ \cite[Fig.\,3]{RyabRCD}\\$\beta_5,\beta_6$ \cite[Fig.\,1]{Mor} \end{tabular}\\

\hline
\begin{tabular}{c}$a_{12}$\\($ \la^* < \la <+\infty$) \end{tabular} &{2}& {\begin{tabular}{c}($-\;-$),($-\;-$)\end{tabular}} & No/Yes &$2A_-$& \begin{tabular}{l} $d_2, d_3$ \cite[Fig.\,5]{RyabRCD}\\$\alpha_3,\alpha_8, \alpha_9,\alpha_{10}$\\ \cite[Fig.\,1]{Mor} \end{tabular}\\

\hline

\end{tabular}
\end{table}

Let us turn to the subsystems $\mathcal{M}_{2,3}$. The analytical solution is found in \cite{EIPVHDan,PVMtt71}, where this set of trajectories is divided into two classes basing on another principle (in fact, by the curve ${\Delta_0}$). The equations for the subsystems can be written in the following algebraic form:
\begin{equation}\notag
\begin{array}{l}
  \displaystyle{\omega_1=-\frac{\ell} {s}- \frac{2 \varkappa \rho  \zeta }{1+\zeta ^2},} \quad  \displaystyle{\omega_2=-\frac{1}{\sqrt{2}(1+\zeta ^2)} \, \sqrt{\frac{\rho^2}{\varkappa s}Z(\zeta )},}\quad  \displaystyle{\omega_3 = \la+2 \varkappa \frac{1-\zeta ^2}{1+\zeta ^2},}\\[2mm]
  \displaystyle{\alpha_1=\frac{\la s(1-\zeta ^4)+2\ell \rho\zeta (1+\zeta ^2)-8\varkappa^3 \zeta ^2}{\varkappa(1+\zeta ^2)^2},} \quad  \displaystyle{\alpha_2=-\frac{2\sqrt{2}\varkappa  }{(1+\zeta ^2)^2} \, \sqrt{\frac{\zeta^2}{\varkappa s}Z(\zeta )},}\\[2mm]
  \displaystyle{\alpha_3 = \frac{\ell (1-\zeta ^2)-2\la \rho s \zeta  }{\varkappa(1+\zeta ^2)}}.
\end{array}
\end{equation}
Here,
\begin{equation}\notag
\begin{array}{l}
  \varkappa^2 = \ell^2+\la^2 s^2, \qquad \rho^2=\displaystyle{1-\frac{2\varkappa^2}{s}},\qquad \zeta = \left\{ \begin{array}{ll} z, & \rho^2 \geqslant 0 \\
                                \ri z, & \rho^2 < 0
                  \end{array}\right., \qquad z \in \mathbb{R},\\[2mm]
  \displaystyle{Z(\zeta )=(\varkappa -2\la s^2) \zeta ^4 +4\ell \rho s \zeta  (1+\zeta ^2)+2\varkappa (1-4\varkappa^2s)\zeta ^2+(\varkappa+2\la s^2).}
\end{array}
\end{equation}
The dynamics is defined by the equation
\begin{equation}\label{eq3_18}
\begin{array}{l}
    \displaystyle{\frac{d\zeta}{dt}=\frac{1}{2\sqrt{2}}  \, \sqrt{\frac{1}{\varkappa s}Z(\zeta)}.}
\end{array}
\end{equation}
Suppose that $s,\ell$ correspond to a level in $\mathcal{M}_{2,3}$ containing no critical points of rank~0. Then, obviously, the number of critical circles in the subsystems $\mathcal{M}_{2,3}$ for the given values $s,\ell$ is equal to the number of real trajectories of the corresponding equation~\eqref{eq3_18} in the phase space $\{(z,{\dot z})\}$ including, of course, the point $z=\infty$.

Thus, to define the number of critical circles we must use the following rule. Make the necessary substitution in equation~\eqref{eq3_18} to the real variable $z$ and consider the obtained polynomial under the radical in the right part. If this polynomial has $2m$ real roots, then the given level of $s,\ell$ in $\mathcal{M}_{2,3}$ contains $m$ critical circles, except for the case when $m=0$ and the highest coefficient is positive. In the latter case we have two critical circles ($z$ ranges over the whole $\overline{\mathbb{R}}$); on each circle the variable $\omega_2$ is of constant sign. Let us emphasize this case as a {\it special} one:
\begin{equation}\label{eq3_19}
\begin{array}{l}
    \rho^2>0, \qquad \varkappa -2\la s^2 > 0, \qquad Z(z)> 0 \quad \forall z \in \mathbb{R}.
\end{array}
\end{equation}

The type of a critical point of rank~1 in the subsystems $\mathcal{M}_{2,3}$ is defined by the symplectic operator generated at this point by the function $F_2=K+(2\la^2-1/s)H$. The eigenvalues of this operator are $\pm \sqrt{m_2}$  \cite{Udgu1}, where
\begin{equation}\notag
\begin{array}{l}
\displaystyle{m_2=-\frac{1}{s^3}\left[2 s^2 - 2 (h + \frac{\la^2}{2}) s+1 \right] \left(8\la^2 s^3-1 \right) = \frac{2}{s^2} (2 \la^2 s^2-s +2 \ell^2)( 8 \la^2 s^3-1)}.
\end{array}
\end{equation}
The first polynomial factor vanishes on the already known curve ${\Delta_0}$ of the tangency between the surfaces $\Pi_1$ and $\Pi_3.$ The zeros of the second factor $8 \la^2 s^3-1=0$ give one more set of degenerate points of rank~1 in the subsystem ${\mathcal{M}_3}$. We denote this set by ${\Delta_3}$. Its image is a part of the cuspidal edge of the surface~$\Pi_3$. Using the motion existence conditions found in \cite{mtt40,Udgu2}, for the integral constants $s,\ell,h$ we have
\begin{equation}\label{eq3_21}
    \begin{array}{l}
      {\Delta_3}:  \quad  \displaystyle{h=h^*+2\ell^2,} \quad
      \left\{
      \begin{array}{ll}  \ell \in \mathbb{R}, & \la \leqslant \la^*\\
                      |\ell| \geqslant  \ell^*, & \la > \la^*
      \end{array}
      \right. , \qquad s=\displaystyle{\frac{1}{2\la^{2/3}}},
    \end{array}
\end{equation}
where
\begin{equation}\notag
    h^*=\frac{1}{2}\la^{2/3}\left(3-\la^{4/3}\right), \quad \ell^*=\displaystyle{\frac{2\la^{2/3}-\sqrt{4+\la^{4/3}}}{\sqrt{2}(\sqrt{4+\la^{4/3}}-\la^{2/3} )^{1/2}}}>0  \qquad (\la > \la^*).
\end{equation}

The explicit analytical formulas for a pair of Morse\,--\,Bott characteristic values in the subsystems $\mathcal{M}_{2,3}$ are too huge. Nevertheless, it is possible to obtain quite simple expressions for calculating these values. Moreover, it is possible to explicitly point out the vector transversal to a hyperbolic trajectory in the direction of which the eight line curve never breaks, except for the special case \eqref{eq3_19}. This makes possible to determine the orientation of atoms of the type $B$ along the direction of increasing the integral $K$.

Consider a case differing from \eqref{eq3_19}. We look for a transversal section to the critical circle in question (at any conveniently chosen critical point of rank~1 belonging to this circle) as an orthogonal complement to the span of $\nabla\Gamma$, $\nabla L$, $\nabla  H$, $\mathop{\rm sgrad}\nolimits  H$.
On each trajectory the variable $z$ oscillates between the roots of the corresponding polynomial $Z(z)$, naturally including the possibility to cross the infinity.
On such a trajectory, let us take the point $x_0$ such that $Z(z)=0$. Then the three gradient vectors are orthogonal to the plane $O\omega_2 \alpha_2$, and the vector $\mathop{\rm sgrad}\nolimits H$ lies in this plane and up to a nonzero multiple has the form
\begin{equation}\notag
\begin{array}{l}
  \bigl(0,\,1,\,0,\,0,\,b,\,0\bigr), \qquad b=\displaystyle{\frac{4\varkappa \, \zeta }{\rho(1+\zeta^2)}}.
\end{array}
\end{equation}
Therefore, for the first vector tangent to a transversal section we take $v_1=(0,\,-b,\,0,\,0,\,1,\,0).$ After this we can take for $v_2$ any nonzero vector orthogonal to five vectors $\nabla  \Gamma$, $\nabla  L$, $\nabla  H$, $\mathop{\rm sgrad}\nolimits H$ and~$v_1.$

The conditional extremum of the function $K$ on a common level of the functions $\Gamma, L, H$ in $\mathbb{R}^6({\boldsymbol\omega},{\boldsymbol\alpha})$ is a critical point of the function with Lagrange multipliers
\begin{equation}\notag
\displaystyle{K_2=K+(2 \la^2 - \frac{1}{s}) H  + 2 s L^2-\frac{2 \varkappa ^2}{s} \Gamma}.
\end{equation}
Obviously, the part of this function that does not contain $L,\Gamma$ coincides with $F_2$. The matrix of the restriction of the quadratic form $d^2 K_2$ on the transversal section calculated in the basis $\{v_1,v_2\}$ turns out to be diagonal; the first Morse\,--\,Bott characteristic value is
\begin{equation}\notag
\begin{array}{l}
    \mu_1=(d^2 K_2) v_1 \cdot v_1=\displaystyle{2\frac{\left[16\varkappa^2\zeta^2+\rho^2(1+\zeta^2)^2\right]^2}{\rho^2(1+\zeta^2)^4}}.
\end{array}
\end{equation}
In particular, its sign is the same on critical circles of the same integral level. Then it is also true for the the value $\mu_2$, since the type of all critical points on these  circles is the same. It is important to emphasize the following. Analyzing the projections of integral manifolds onto the plane $O\omega_1 \omega_2$
in the neighborhood of the subsystems $\mathcal{M}_{2,3}$ we see that, similar to the case of the subsystem ${\mathcal{M}_1},$ the critical surface of any hyperbolic circle never breaks in the direction of the axis $O\omega_2$. This means that for the atoms $B$ the vector $v_1$ always points to the outer part of the eight line curve. It then follows that if $\mu_1>0$ the function $K$ increases to the head of the atom (the atom has its head up), and if $\mu_1<0$ the function $K$ decreases to the head of the atom (the atom has its head down). Thus, the direction of the edges for non-symmetric atoms is defined by the sign of $\rho^2.$

\begin{figure}[htp]
\centering
\includegraphics[width=60mm,keepaspectratio]{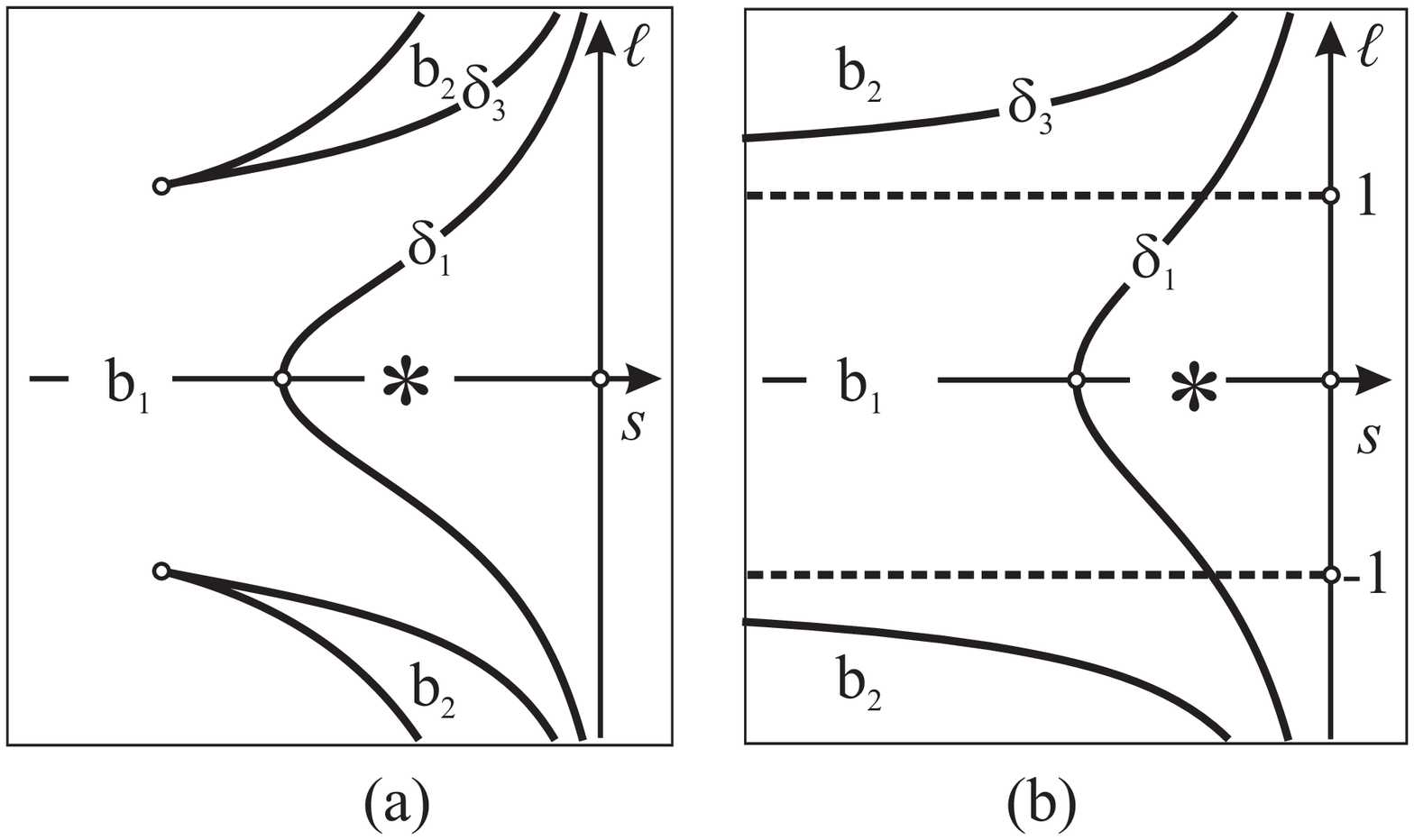}\\
\caption{The diagram of the subsystem ${\mathcal{M}_2}$: (a) $\la>0$; (b) the limit case $\la=0$}\label{fig_sys2}
\end{figure}

For ${\mathcal{M}_2}$ by definition we have $s<0$, so $\rho^2>0$, $m_2<0$ and $\mu_1>0$. In $\mathcal{M}_2$ there are no degenerate points. The $(S,L)$-diagram is the same as the bifurcation diagram in \cite{Udgu2} and consists of the sets $\delta_1, \delta_3$. No transformations of the diagram occur for $\la>0$. All critical points of rank~1 have the elliptic type. As the integral $K$ increases, on any iso-energetic level ${Q_{\ell,h}^3}$ any critical circle of ${\mathcal{M}_2}$ produces the bifurcation $A_+$ of the torus birth.
In Fig.~\ref{fig_sys2}, together with the $(S,L)$-diagram of the second critical subsystem we show the admissible region including the domain $b_1$ with one critical circle and two symmetric with respect to $\ell=0$ domains $b_2$ with two critical circles. In the domain marked with an asterisk no motions exist. The properties of the corresponding levels and atoms are collected in Table~\ref{table3}. We can see that no new features appear in the subsystem ${\mathcal{M}_2}$ compared to the classical problem.

\begin{table}[ht]
\centering
\small

\begin{tabular}{|c|c| c| c| c|c|}
\multicolumn{6}{r}{{Table \myt\label{table3}}}\\
\hline
\begin{tabular}{c}{Domain}\\[-3pt]{(life time)} \end{tabular}
&\begin{tabular}{c}{Critical}\\[-3pt]{circles}\end{tabular}&\begin{tabular}{c}{Morse\,--\,Bott}\\[-3pt]{char. vals.}\end{tabular}
&\begin{tabular}{c}{Exit to }\\[-3pt]{$\la=0$/$\ell=0$}\end{tabular}
&\begin{tabular}{c}{Atom}\end{tabular} &\begin{tabular}{c}{Analogues}\end{tabular}\\

\hline
\begin{tabular}{c}$b_1$\\($ 0 \leqslant \la <+\infty$) \end{tabular} &{1}& {($+\;+$)} & Yes/Yes &$A_+$ & \begin{tabular}{l} 1 \cite[Fig.\,6.3]{KhBook88}\\ $a_1$ \cite[Fig.\,2]{RyabRCD}\\$\alpha_1$ \cite[Fig.\,11]{BRF}\\$\alpha_1$ \cite[Fig.\,1]{Mor} \end{tabular}\\
\hline
\begin{tabular}{c}$b_2$\\($ 0 \leqslant \la <+\infty$) \end{tabular} &{2}& {\begin{tabular}{c}($+\;+$),($+\;+$)\end{tabular}} & Yes/No &$2A_+$ &
\begin{tabular}{l} Transit {III}$\to${VI}\\ \cite[Fig.\,6.1{\it d}]{KhBook88}\\$\alpha_2$ \cite[Fig.\,11]{BRF} \end{tabular}\\
\hline
\end{tabular}
\end{table}

It is known from the investigations of the classical Kowalevski problem \cite{KhPMM83} that in the subsystem ${\mathcal{M}_3}$ the atoms $C_2$ appear. For the non-special case, as shown above, any two critical circles on the same integral level have the same distribution of the signs in the pairs of the Morse\,--\,Bott characteristic values.
Consider the special case \eqref{eq3_19}. On the $(s,\ell)$-plane, $\rho^2$ and $\varkappa -2\la s^2$ are positive in the domain $\la^2 s^2(2s^2-1)<\ell^2<s(1-2\la^2s)/2$.
This domain lies completely inside the closed curve $\Delta_0$ and, in the case $\la<1$, is restricted at the right side by the branch of the curve $\delta_2$. The intersection of this domain with the axis $\ell=0$ is the interval $s \in (0,\min \{1/2,1/2\la^2\}).$ Since $Z>0$ for all $z\in \bR$ we can choose $z=0$ for simplicity. Then the vectors defining the transversal section to the periodic solution are easily found:
$$
\begin{array}{l}
  v_1=\displaystyle{\left(-(1+2\la^2s),\,0,\,0,\,0,\,\frac{\sqrt{2s(1-2\la^2s)}}{\sqrt{1+2s}}[(1+s)\la^2s-1],\,\la s(3-2\la^2s)\right),} \\[3mm]
  v_2=\displaystyle{\left(0,\,\frac{\la \sqrt{s(1+2s)}}{\sqrt{2(1-2\la^2s)}},\,1,\,0,\,0,\,0 \right).}
\end{array}
$$
The eigenvalues of the matrix of the quadratic form $d^2 K_2$ calculated at this pair of vectors are
$$
\begin{array}{l}
  \mu_1=\displaystyle{\frac{4s}{1+2s}[1+\la^2s^2(5-2\la^2s)]^2}, \quad   \mu_2=\displaystyle{\frac{1}{s}(8\la^2s^3-1)}.
\end{array}
$$
Obviously, under the conditions in question we obtain $\mu_1>0,$ $\mu_2<0$. Thus, the distribution of the signs is the same for both trajectories at this integral level. Namely, for the chosen basis it is $(+,-)$. So we have to state that both calculations of the points type and the Morse\,--\,Bott characteristic values based on the local analysis
do not provide a way to differ between the atoms $2B$ and $C_2$. Nevertheless, in the enhanced space the considered domain has an exit to the classical analogue $\la=0$, and the comparison with the results of \cite{KhPMM83,KhBook88} for the classical case proves that the special case corresponds to the atom $C_2$. Finally we come to the following statement.

\begin{pre}\label{propos6}
{\it As the integral $K$ increases on an iso-energetic level ${Q_{\ell,h}^3}$, we obtain the following bifurcations at the points of the critical subsystem ${\mathcal{M}_3}$ belonging to non-degenerate critical circles{\rm :}

{\rm 1)} for elliptic trajectories we have the atom $A_+$ if $\rho^2>0$ and the atom $A_-$ if $\rho^2<0;$

{\rm 2)} for one hyperbolic trajectory on a critical level of $K$ we have the atom $B_-$ if $\rho^2>0$ and the atom $B_+$ if $\rho^2 < 0;$

{\rm 3)} for two hyperbolic trajectories on a critical level of $K$ we have two atoms $B_-$ if $\rho^2 > 0$ and two atoms $B_+$ if $\rho^2 < 0,$ except for the special case~\eqref{eq3_19}, in which we have the atom $C_2.$}
\end{pre}

\begin{figure}[htp]
\centering
\includegraphics[width=150mm,keepaspectratio]{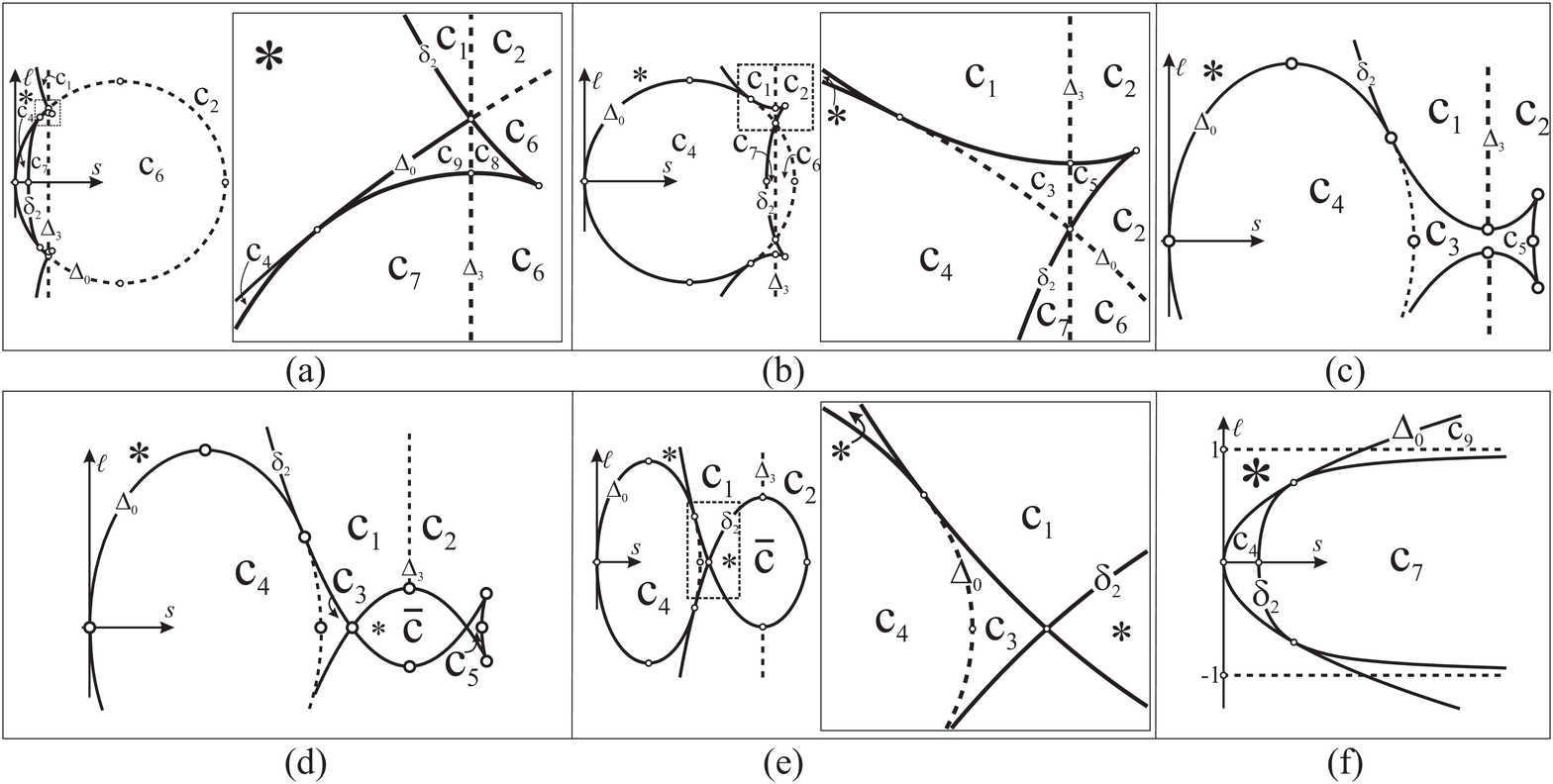}\\
\caption{The diagram of the subsystem ${\mathcal{M}_3}$}\label{fig_sys3}
\end{figure}

The $(S,L)$-diagram of the critical subsystem ${\mathcal{M}_3}$ is obtained from the bifurcation diagram given in \cite{Udgu2} by adding the sets $\Delta_0$ and $\Delta_3$, where the latter is the image on the $(s,\ell)$-plane of the curve \eqref{eq3_21}. The admissible region does not include the following components of the complement to the diagram: the domain adjacent to the axis $s=0$ and bounded by the branches of the curves ${\Delta_0}, \delta_2$ for all $\la$;
the domain bounded by the curve $\delta_2$ between the two points of its intersection with the axis $\ell=0$ with $r \neq 0$ for $\la>\la^*$.
The transformations of the diagram for $\la>0$ take place at the same values of the parameter as in the subsystem ${\mathcal{M}_1}$.
The types of the diagram of the subsystem ${\mathcal{M}_3}$ are shown in Fig.~\ref{fig_sys3}: $(a)$~$0<\la<\la_*$; $(b)$~$\la_*<\la<1$; $(c)$~$1<\la<\la^*$; $(d)$~$\la^*<\la<\sqrt{2}$; $(e)$~$\la>\sqrt{2}$; $(f)$~is the limit case $\la=0$. Dashed lines show the curves of degenerate points ${\Delta_0},{\Delta_3}$ except for that part of the curve ${\Delta_0}$ which is the outer boundary of the admissible region; this part is drawn with a solid line. Similar to the previous cases, the asterisk stays for the domains that has no critical motions. Note that the absence of critical motions in the domain denoted by $\overline{c}$ (Fig.~\ref{fig_sys3},\,{\it d,e}) was first proved in \cite{RyabDis} for the points on the axis~$\ell=0$.

Applying the obtained results to the points of the domains $c_1-c_9$ in the image of the subsystem ${\mathcal{M}_3}$ on the $(s,\ell)$-plane leads to the information about the characteristics and atoms collected in Table~\ref{table4}.
We see that all domains except for $c_1$ and $c_8$ have common points with the correspondent domains in the investigated earlier problems ($\la=0$ or $\ell=0$) when the enhanced diagram in the $(s,\ell,\la)$-space is considered. Therefore the only additional information for the atoms here is their orientation. In particular, the existence of the atom $C_2$ in the domain $c_4$ and of two atoms $B$ in the domain $c_9$ is proved in the works \cite{KhPMM83, KhDan83, KhBook88} (in different notation). The fact that the domain $c_5$ corresponds to two atoms $B$ follows from the results of \cite{RyabRCD}.
In the new domains $c_1$ and $c_8$, as it is proved above, critical circles are of elliptic type, the number of circles is calculated according to the given criteria, and the orientation of the atoms is defined by the Morse\,--\,Bott characteristic values.

\clearpage

\begin{table}[ht]
\centering
\small
\begin{tabular}{|c|c| c| c| c|c|}
\multicolumn{6}{r}{{Table \myt\label{table4}}}\\
\hline
\begin{tabular}{c}{Domain}\\[-3pt]{(life time)} \end{tabular}
&\begin{tabular}{c}{Critical}\\[-3pt]{circles}\end{tabular}&\begin{tabular}{c}{Morse\,--\,Bott}\\[-3pt]{char. vals.}\end{tabular}
&\begin{tabular}{c}{Exit to }\\[-3pt]{$\la=0$/$\ell=0$}\end{tabular}
&\begin{tabular}{c}{Atom}\end{tabular} &\begin{tabular}{c}{Analogues}\end{tabular}\\
\hline
\begin{tabular}{c}$c_1$\\($ 0 < \la <+\infty$) \end{tabular} &{1}& {($-\;-$)} & No/No &$A_-$ & {}\\

\hline
\begin{tabular}{c}$c_2$\\($ 0 < \la <+\infty$) \end{tabular} &{1}& {($-\;+$)} & No/Yes &$B_+$ & \begin{tabular}{l} $a_5$ \cite[Fig.\,2]{RyabRCD}\\$\beta_3$ \cite[Fig.\,1]{Mor}\end{tabular}\\

\hline
\begin{tabular}{c}$c_3$\\($ \la_* < \la < +\infty$) \end{tabular} &{2}& {\begin{tabular}{c}($-\;-$),($-\;-$)\end{tabular}} & No/Yes &$2A_-$ & \begin{tabular}{l} $b_4$ \cite[Fig.\,3]{RyabRCD}\\$\alpha_7$ \cite[Fig.\,1]{Mor}\end{tabular}\\

\hline
\begin{tabular}{c}$c_4$\\($ 0 \leqslant \la < +\infty $) \end{tabular} &{2}& {\begin{tabular}{c}($+\;-$),($+\;-$)\end{tabular}} & Yes/Yes &$C_2$ & \begin{tabular}{l} 8 \cite[Fig.\,6.3]{KhBook88}\\ $a_4, b_5$ \cite[Fig.\,2,3]{RyabRCD}\\$\beta_2$ \cite[Fig.\,11]{BRF}\\$\gamma$ \cite[Fig.\,1]{Mor} \end{tabular}\\

\hline
\begin{tabular}{c}$c_5$\\($ \la_* < \la <\sqrt{2} $) \end{tabular} &{2}& {\begin{tabular}{c}($-\;+$),($-\;+$)\end{tabular}} & No/Yes &$2B_+$& \begin{tabular}{l} $b_3$ \cite[Fig.\,3]{RyabRCD}\\$\beta_4$ \cite[Fig.\,1]{Mor} \end{tabular}\\

\hline
\begin{tabular}{c}$c_6$\\($ 0 < \la < 1$) \end{tabular} &{1}& {($+\;+$)} & No/Yes &$A_+$& \begin{tabular}{l} $a_3, a_4$ \cite[Fig.\,2]{RyabRCD}\\$\alpha_4$ \cite[Fig.\,1]{Mor}\end{tabular}\\

\hline
\begin{tabular}{c}$c_7$\\($ 0 \leqslant \la < 1$) \end{tabular} &{1}& {($+\;-$)} & Yes/Yes &$B_-$& \begin{tabular}{l} 7 \cite[Fig.\,6.3]{KhBook88}\\$a_3$ \cite[Fig.\,2]{RyabRCD}\\$\beta_1$ \cite[Fig.\,11]{BRF}\\$\beta_2$ \cite[Fig.\,1]{Mor}\end{tabular}\\

\hline
\begin{tabular}{c}$c_8$\\($ 0 < \la < \la_*$) \end{tabular} &{2}& {\begin{tabular}{c}($+\;+$),($+\;+$)\end{tabular}} & No/No &$2A_+$& {} \\

\hline
\begin{tabular}{c}$c_9$\\($ 0\leqslant \la < \la_*$) \end{tabular} &{2}& {\begin{tabular}{c}($+\;-$),($+\;-$)\end{tabular}} & Yes/No &$2B_-$& \begin{tabular}{l} E\,\cite[Fig.\,2]{KhDan83}\\$\beta_3$ \cite[Fig.\,11]{BRF} \end{tabular}\\

\hline

\end{tabular}
\end{table}

\begin{flushleft}
{\bf{\S\,3. The Smale\,--\,Fomenko diagrams}}
\end{flushleft}
It follows from the definition of an iso-energetic topological invariant \cite{FomFAN88} that the set of parameters separating different Fomenko graphs on ${Q_{\ell,h}^3}(\la)$ must include, in addition to Smale's diagram, the image of the set of degenerate critical points of rank~1 because crossing such points causes transformations in the graph occurring without changing the topology of ${Q_{\ell,h}^3}(\la)$. Then the set of curves separating non-equivalent graphs consists of the curves $\delta_j$ $(j=1,2,3)$ given by equations \eqref{eq2_7} and of the curves ${\Delta_0},{\Delta_1},{\Delta_3}$ defined according to \eqref{eq3_5}, \eqref{eq3_6}, \eqref{eq3_21}. The equations of these six curves (without any investigation of restrictions on the coordinates or parameters) treated as a separating set were first obtained in \cite{GashMttTop}.

\begin{df}\label{def13}
The union of ${\Sigma_{LH}}$ with the image of all degenerate points of rank $1$ under the map $L{\times}H$ is called the Smale\,--\,Fomenko diagram and is denoted by ${\Sigma'_{LH}}$.
\end{df}

Such diagrams for the classical problems of the rigid body dynamics are constructed in the works by A.A.\,Oshemkov \cite{OshEng,OshRus}.
Note that in this problem we find quite a short list of basic atoms. If more complicated atoms appear, the transformations of molecules can take place on the same iso-energetic manifold without crossing degenerate points (see, for example, the results of numerical modeling in the work \cite{Mosk}).
Nevertheless, to-day we cannot say whether it is possible, using only the local analysis of singularities, to predict such transformations and add the corresponding separating set to the Smale\,--\,Fomenko diagram.

\begin{teo}\label{th8}
{\it In the Kowalevski\,--\,Yehia problem there exist ten types of the Smale\,--\,Fomenko diagrams ${\Sigma'_{LH}}(\la)$ stable with respect to the parameter $\la$. The separating values of the parameter are $0$, $\la_1$, $\la_3$, $\la_*$, $1$, $\la^*$, $\la_4$, $2\sqrt{\sqrt{2}-1}$, $\la_2$, $\sqrt{2}$, where}
\begin{equation*}
  \la_3 = \left( \frac{45}{2^{1/3}}-\frac{99}{4 \cdot 2^{2/3}}-\frac{161}{8} \right)^{1/4} \approx 0.0287; \quad \la_4 = \left[\frac{1}{2}(3-\sqrt[3]{2})^3\right]^{1/4} \approx 1.2740.
\end{equation*}
{\it The enhanced diagram in the $(\ell,h,\la)$-space generates the additional division of Smale's chambers $\mtA-\mtH$ into $29$ chambers. The chambers $\mtC,\mtD,\mtF,\mtG$ are not divided. The other ones have the following division: $\mtA_{1}-\mtA_{13}$, $\mtB_{1}-\mtB_{3}$, $\mtE_{1}-\mtE_{6}$, $\mtH_{1}-\mtH_{3}$. Inside each of the arising $29$ chambers the Fomenko graph is preserved.}
\end{teo}
\doc. The intersection of the curves ${\Delta_j}$ with the main diagram consists of degenerate points of rank $0$, i.e., of the singular points of Smale's diagram. Therefore, all possible transformations of the sets ${\Delta_j} \cap {\Sigma_{LH}}$ with respect to $\la$ are already found. To classify the Smale\,--\,Fomenko diagrams we need to add the values of $\la$ at which the set ${\Delta_0}\cup {\Delta_1}\cup {\Delta_3}$ is restructured. Obviously, there are no triple intersections in this set. The self-intersection in the $(\ell,h)$-plane arises only on the curve ${\Delta_0}$ at $\la \leqslant \la_*$. Writing down the conditions for the intersections ${\Delta_0} \cap {\Delta_1}$ and ${\Delta_0} \cap {\Delta_3}$ does not give new separating cases either. Let us consider the points of intersection ${\Delta_1} \cap {\Delta_3}$. Rewrite \eqref{eq3_6} in the explicit form
\begin{equation}\label{eq4_7}
    {\Delta_1}: \quad \displaystyle{\ell = \pm \frac{2}{3\sqrt{3}}(h-\frac{\la^2}{2})^{3/2}}, \quad \displaystyle{\frac{\la^2}{2}\leqslant h \leqslant h_*}.
\end{equation}
Here $h_*$ corresponds to the boundary value $s_*$ in \eqref{eq3_6}. Ignoring the restrictions, for the solutions of the system of equations \eqref{eq3_21}, \eqref{eq4_7} we have
\begin{eqnarray}
& & h=\frac{1}{2} \left(3 \la^{2/3} + \la^2\right),\quad \ell^2=\frac{\la^2}{2}; \label{eq4_9}\\
& & h=\frac{1}{4} \left[ 2 \la^2 -3 \la^{2/3}+ 3 \sqrt{3(2 - \la^{4/3})}\right], \quad \ell^2=\frac{1}{8}\left[4 \la^2-9 \la^{2/3} + 3 \sqrt{3(2 - \la^{4/3})}\right].\label{eq4_11}
\end{eqnarray}

\begin{figure}[!ht]
\centering
\includegraphics[width=155mm,keepaspectratio]{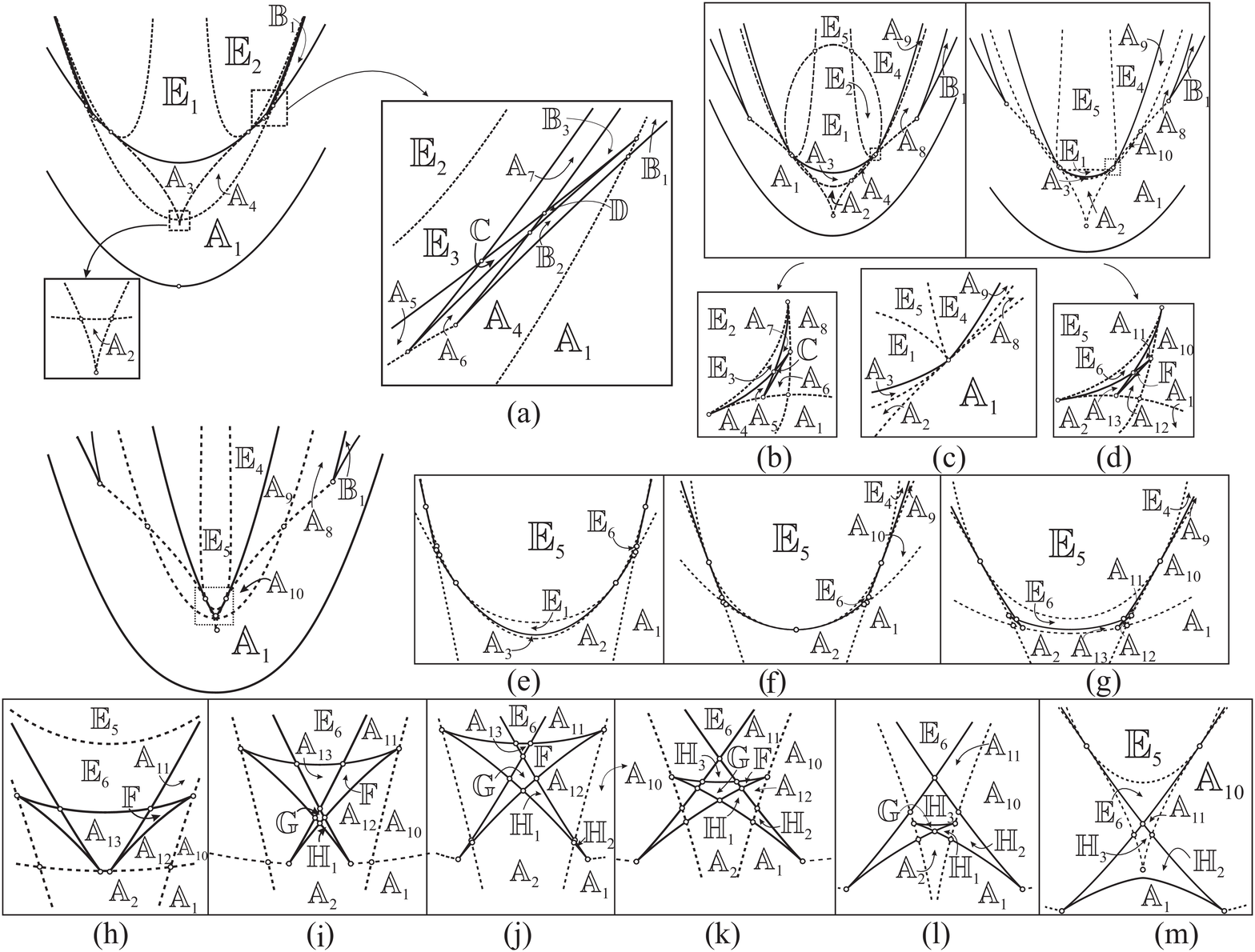}\\
\caption{Chambers of the Smale\,--\,Fomenko diagrams}\label{figsm1}
\end{figure}

The pair of points \eqref{eq4_9} always exists and satisfies all inequalities in \eqref{eq3_21} and \eqref{eq4_7}. Consider the solution \eqref{eq4_11} real for all $\la<2^{3/4}.$ First, we check the conditions on $h$ on the curve ${\Delta_1}.$ In~\eqref{eq4_11}, put $h=h_*$ to obtain the equation
\begin{equation}\notag
    1+48 \la^{4/3}+12 \la^{8/3}-8 \la^{4}-6\sqrt{3(2-\la^{4/3})}(\la^{2/3}+4 \la^{2})=0.
\end{equation}
The latter has the root $\la_*$ of degree 3 and the simple root $\la_3$ in the interval $(\la_1,\la_*)$. The condition $h \leqslant h_*$ holds for $\la \geqslant \la_3$. The root $\la_*$ does not affect this inequality; this root is connected with a singular point on the curve $\delta_2$. Now let us check the conditions for $h$ on the curve ${\Delta_3}$ at the points \eqref{eq4_11}. We have
$$
h-h^*=\frac{1}{4}\left[\la^2-9\la^{2/3}+3\sqrt{3(2-\la^{4/3})} \right].
$$
This value is non-negative when $\la \leqslant (3/2)^{3/4} \approx 1.3554$ and, consequently, when ${\la \leqslant \la^*}$.
Let $\la>\la^*$. Then the value
$$
\ell^2 -(\ell^*)^2 = \frac{1}{8} \left[ 3 \la^{2/3} (1+\la^{4/3} )+3 \sqrt{3} \sqrt{2-\la^{4/3}}- (4 + \la^{4/3})^{3/2}\right]
$$
must be non-negative. The right hand part has the root $\la_*$ of degree 3 not affecting the sign, and the simple root $\la_4$. Thus, the pair of the intersection points of the curves ${\Delta_1}$ and ${\Delta_3}$ defined by equations \eqref{eq4_11} exists if $\la_3 \leqslant \la \leqslant \la_4$. All separating values are found. Building the correspondent diagrams we reveal all new chambers. \hfill $\square$

On Fig.~\ref{figsm1}, the illustrations are given showing all necessary details of the Smale\,--\,Fomenko diagrams and the notation of the chambers. The case $(a)$ corresponds to the values $\la<\la_1$. Figures $(b)$--$(d)$ show crossing the value $\la_*$, disappearing of the chambers $\mtA_4 - \mtA_7, \mtC,\mtE_2,\mtE_3$ and birth of new chambers $\mtA_{10} - \mtA_{13}, \mtE_6, \mtF$.  Figures $(e)$--$(g)$ show crossing the value $\la=1$, disappearing of the chambers $\mtA_3$ and $\mtE_1$. The rest of the cases are $(h)$~$1<\la<\la^*$; $(i)$~$\la^*<\la<\la_4$; $(j)$~$\la_4<\la<\la_5$; $(k)$~$\la_5<\la<\la_2$; $(l)$~$\la_2<\la<\sqrt{2}$; $(m)$~$\la>\sqrt{2}$.

Note that in the work \cite{GashDis} we also find the statement on 29 chambers and the table containing the corresponding coordinates for the example points in the $(\ell,h)$-plane. Nevertheless, we could not find any restrictions connected with the existence conditions for degenerate critical points of rank~1 (critical circles) on the curves ${\Delta_1},{\Delta_3}$ given here in \eqref{eq3_6}, \eqref{eq3_21}. Without using these conditions, the number of chambers with non-empty integral manifolds should be equal to 31. In the work \cite{GashMttTop} we read: ``We define the number of critical circles on each critical level of the integral $K$ using the relations of P.V.\,Kharlamov \cite{PVMtt71,Gash4}''. Nevertheless, in the work \cite{Gash4} this question is not discussed, and it does not seem possible to extract any existence conditions directly from the works \cite{PVLect,EIPVHDan,PVMtt71} without any additional analytical transformation (for example, presenting the solutions in some algebraic form as it was done above, with further detailed analysis of the obtained polynomials). It would be interesting to find out the motives for the conclusions of \cite{GashDis} and for the corresponding classification of the Smale\,--\,Fomenko chambers (the work \cite{GashDis} in the part dealing with the Kowalevski\,--\,Yehia case is free to access at \verb"http://iamm.ac.donetsk.ua/upload/iblock/c60/chapter_9.pdf").

\begin{flushleft}
{\bf{\S\,4. Classification of the Fomenko graphs}}
\end{flushleft}
Having obtained all information on the existence conditions and classes of critical points in the critical subsystems, we can now build all \textit{typical} Fomenko graphs applying the same algorithm. Let us fix the value $\la$ and choose a typical point $(\ell,h)$ in one of the Smale\,--\,Fomenko chambers. Let $\tau_{\ell,h}$ denote the corresponding straight line parallel to the axis $Ok$. It is easy to see that its intersection with the admissible region $J(P^5) \subset \mathbb{R}^3{(\ell,h,k)}$ is an effectively calculated segment $[k_{\min},k_{\max}]$. Indeed, the value $k_{\min}$ is obtained in the single point of intersection with the image of the subsystem ${\mathcal{M}_2}$, which on the $(S,L)$-diagram corresponds to the unique negative root of the equation in $s$
\begin{equation}\label{nq5_1}
    s^3-(h-\frac{\la^2}{2})s^2+\ell^2=0.
\end{equation}
The value $k_{\max}$ is obtained in the single point of intersection with the image of the subsystem ${\mathcal{M}_1}$ in one of the domains $a_1$ or $a_{12}$, which on the $(S,L)$-diagram corresponds to the unique negative root of the equation in $s$
\begin{equation}\label{nq5_2}
    4\la^2 s^3-2(h+\frac{\la^2}{2}-2\ell^2)s+1=0.
\end{equation}

\begin{figure}[ht]
\centering
\includegraphics[width=120mm,keepaspectratio]{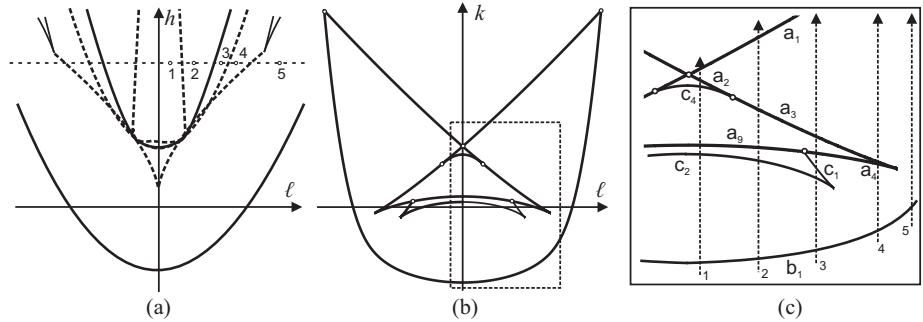}\\
\caption{The diagrams and the paths for the Fomenko graphs $(\la=0.8, h=2.5)$}\label{fig_smalefomd}
\end{figure}

\begin{figure}[!ht]
\centering
\includegraphics[width=150mm,keepaspectratio]{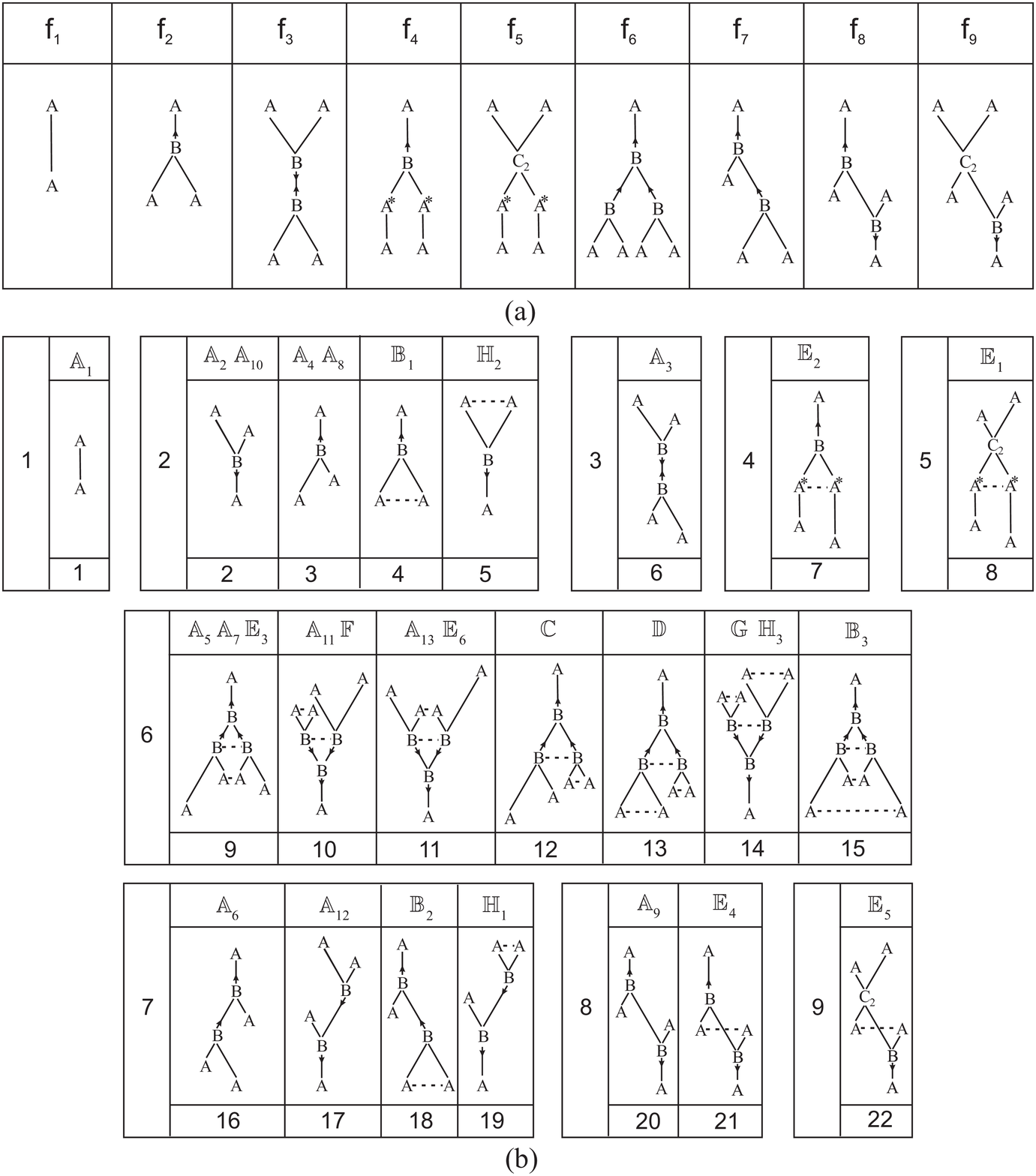}\\
\caption{The groups of graphs and the Fomenko graphs}\label{fig_fomgr}
\end{figure}

All other values of $s$ at the points of intersection with the subsystems ${\mathcal{M}}_j$ are also found from equations \eqref{nq5_1}, \eqref{nq5_2}. We see that for all three subsystems we obtain not more than six points. For each of them the above tables give the corresponding atom. The critical values of $K$ obtained from the equations of the bifurcation surfaces $\Pi_j$ should be sorted in increasing order. This way, we get the whole sequence of atoms in the Fomenko graph together with their orientation along the $Ok$ axis.  It is convenient to show this procedure on the so-called iso-energetic bifurcation diagrams, i.e., on the cross-sections of the bifurcation diagram $\Sigma$ by the planes $h={\rm const}$, due to the obvious fact that the manifolds $\{x \in P^5:H(x)=h\}$ and their images are compact. These cross-sections are investigated in \cite{mtt40}, where the existence conditions for the critical motions in the subsystems ${\mathcal{M}}_j$ are obtained in terms of the parameter $h$. As an example, let us consider some ``average'' values $\la=0.8$ and $h=2.5$. In Fig.~\ref{fig_smalefomd},{\it a} the corresponding Smale\,--\,Fomenko diagram is shown. The given level of $h$, as $\ell$ increases from zero, crosses five chambers $\mtE_5,\mtE_4,\mtA_9,\mtA_8,\mtA_1$ (the $\ell$-values in these chambers are marked by the numbers 1,\dots,5). In the $h$-section of the diagram $\Sigma(\la)$ (see Fig.~\ref{fig_smalefomd},{\it b}) the Fomenko graphs are defined with the bifurcations occurring along the lines $\ell={\rm const}$ as $k$ increases (five dashed arrows in Fig.~\ref{fig_smalefomd},{\it c}):
$$
\begin{array}{lllll}
{1)}&{\mtE_5:}&{b_1\to c_2\to a_9\to c_4\to a_2\to a_1}&\Leftrightarrow&
{A_+ \to B_+ \to (A_+,A_-) \to C_2 \to A_- \to A_-}\\
{2)}&{\mtE_4:}&
{b_1\to c_2\to a_9\to a_3\to a_1} &\Leftrightarrow& {A_+\to B_+ \to (A_+,A_-)\to  B_-\to A_-}\\
{3)}&{\mtA_9:}&
{b_1\to c_2\to c_1\to a_4\to a_3\to a_1}
&\Leftrightarrow& {A_+\to B_+ \to A_- \to A_+\to  B_-\to A_-}\\
{4)}&{\mtA_8:}&
{b_1\to a_4\to a_3\to a_1}
&\Leftrightarrow& {A_+ \to A_+ \to B_- \to A_-}\\
{5)}&{\mtA_1:}& {b_1\to a_1}&\Leftrightarrow& {A_+ \to A_-}
\end{array}
$$
Fulfilling this procedure for all chambers we get the complete classification of the Fomenko graphs collected in Table~\ref{table5}. Let us give the necessary comments on the terminology and the notation in this table. The column titled ``Graph'' shows the group of the Fomenko graph and (in parentheses) its number. Here we consider the group of the Fomenko graph as a class of identical molecules in the sense of~\cite{BolFom}. The molecule of an iso-energetic manifold ${Q_{\ell,h}^3}$ is, first, the graph itself, i.e., a topological space obtained from ${Q_{\ell,h}^3}$ by identifying with one point all points on each connected component of any integral manifold of the type ${Q_{\ell,h}^3} \cap \{K={\rm const}\};$ the graph vertices are the points obtained from the components containing critical points of $K$ on ${Q_{\ell,h}^3}$. Second, each vertex is supplied with the notation of the atom that arises in the neighborhood of the critical connected component. The question is what molecules should be considered equivalent. In \cite{BolFom}, two molecules are called \textit{identical} if there exists a homeomorphism of the graphs taking edges to edges, vertices to vertices, and this homeomorphism can be extended to the atoms themselves (in the obvious sense, considering atoms as Liouville foliated sets). According to this definition, we obtain nine groups of equivalent Fomenko graphs. They are shown in Fig.~\ref{fig_fomgr},{\it a}. The arrowhead on a graph denotes the direction to the ``head'' of the non-symmetric atom $B$ (i.e., to the outer circle surrounding the eight line curve). The groups $f_1-f_6$ correspond to the graphs of the types $W_1-W_6$ found in the work \cite{GashMttTop}. The groups $f_7-f_9$ are new.  In the groups $f_7,$ $f_8$ the atoms $B$ are not connected ``head to head'' as in the group $f_3$ and the type $W_3$ of \cite{GashMttTop,GashDis}. The group $f_9$ resembles the graph $W_7$ of the works \cite{GashMttTop,GashDis} since these graphs treated purely as topological spaces are homeomorphic. But if we strictly follow the picture of the graph $W_7$ in \cite{GashMttTop,GashDis}, then we see that in $f_9$, not alike $W_7$, the edge from the atom $C_2$ goes to the ``leg'', not to the ``head'', of the atom $B$; therefore, the resulting Liouville foliation is different.

\begin{table}[!ht]
{
\centering
\small
\begin{tabular}{| c| c|c|c|c|c|}
\multicolumn{6}{r}{{Table \myt\label{table5}}}\\
\hline
{Chamber}
&\begin{tabular}{c}{Life time}\\[-3pt]{w.r.t $\la$}\end{tabular}&\begin{tabular}{c}{Exit to}\\[-3pt]{$\la=0$}/{$\ell=0$}
\end{tabular}&\begin{tabular}{c}{Arcs}\\[-3pt]{sequence}\end{tabular}
&\begin{tabular}{c}{Graph}\end{tabular}
&\begin{tabular}{c}{Marked}\\[-3pt]{molecule}\end{tabular}\\
\hline
$\mtA_1$ & $ 0 \leqslant \la <+\infty$ & Yes/Yes &{$b_1\to a_1$}& 1(1) & \begin{tabular}{c}$A$\,\cite[Table~3]{BRF},\\ \,\cite[Table~8]{Mor}\end{tabular}\\

\hline
$\mtA_2$ & $ 0 < \la <+\infty$ & No/Yes &\begin{tabular}{c}{$b_1\to a_6\to a_2\to a_1$}\end{tabular}&2(2) & $B$\,\cite[Table~8]{Mor}\\

\hline
$\mtA_3$ & $ 0 \leqslant \la < 1$ & Yes/Yes &\begin{tabular}{c}{$b_1\to c_6\to c_7\to$}\\[-3pt] {$\to a_6\to a_2\to a_1$}\end{tabular}&3(6)& \begin{tabular}{c}$C$\,\cite[Table~3]{BRF},\\ \,\cite[Table~8]{Mor}\end{tabular}\\

\hline
$\mtA_4$ & $ 0 \leqslant \la <\la_*$ & Yes/No &{$b_1\to c_6\to c_7\to a_1$}&2(3)& $B$\,\cite[Table~3]{BRF}\\

\hline
$\mtA_5$ & $ 0 \leqslant \la <\la_*$ & Yes/No &\begin{tabular}{c}{$b_1\to c_6\to a_8\to$}\\[-3pt] {$\to a_7\to c_7\to a_1$}\end{tabular}&6(9)& $J$\,\cite[Table~3]{BRF}\\

\hline
$\mtA_6$ & $ 0 < \la < \la_*$ & No/No &\begin{tabular}{c}{$b_1\to a_4\to a_3\to$}\\[-3pt] {$\to c_6\to c_7\to a_1$}\end{tabular}&7(16)& \\

\hline
$\mtA_7$ & $ 0 < \la < \la_*$ & No/No &\begin{tabular}{c}{$b_1\to a_4\to c_8\to$}\\[-3pt] {$\to c_9\to a_3\to a_1$}\end{tabular}
&6(9)& \\

\hline
$\mtA_8$ & $ 0 < \la <+\infty$ & No/No &\begin{tabular}{c}{$b_1\to a_4\to a_3\to a_1$}\end{tabular}&2(3)& \\

\hline
$\mtA_9$ & $ 0 < \la <+\infty$ & No/No &\begin{tabular}{c}{$b_1\to c_2\to c_1\to$}\\[-3pt] {$\to a_4\to a_3\to a_1$}\end{tabular}&8(20)& \\

\hline
$\mtA_{10}$ & $ \la > \la_* $ & No/No &\begin{tabular}{c}{$b_1\to c_2\to c_1\to a_1$}\end{tabular}&2(2)& \\

\hline
$\mtA_{11}$ & $ \la > \la_* $ & No/No &\begin{tabular}{c}{$b_1\to c_2\to a_{11}\to$}\\[-3pt] {$\to a_{10}\to c_1\to a_1$}\end{tabular}&6(10)& \\

\hline
$\mtA_{12}$ & $ \la_* < \la < \la_2 $ & No/No &\begin{tabular}{c}{$b_1\to a_6\to a_2\to$}\\[-3pt] {$\to c_2\to c_1\to a_1$}\end{tabular}
&7(17)& \\

\hline
$\mtA_{13}$ & $ \la_* < \la < \la_5 $ & No/Yes &\begin{tabular}{c}{$b_1\to a_6\to c_5\to$}\\[-3pt] {$\to c_3\to a_2\to a_1$}\end{tabular}
&6(11)& $F$\,\cite[Table~8]{Mor}\\

\hline

$\mtB_{1}$ & $ 0 \leqslant \la < +\infty $ & Yes/No &\begin{tabular}{c}{$b_2\to a_3\to a_1$}\end{tabular}&2(4)& $F$\,\cite[Table~3]{BRF}\\

\hline
$\mtB_{2}$ & $ 0 < \la < \la_3 $ & No/No &\begin{tabular}{c}{$b_2\to a_3\to c_6\to$}\\[-3pt] {$\to c_7\to a_1$}\end{tabular}&7(18)& \\

\hline

$\mtB_{3}$ & $ 0 \leqslant \la < \la_1 $ & Yes/No &\begin{tabular}{c}{$b_2\to c_8\to c_9\to$}\\[-3pt] {$\to a_3\to a_1$}\end{tabular}
&6(15)& $G$\,\cite[Table~3]{BRF}\\

\hline
$\mtC$ & $ 0 <\la < \la_* $ & No/No &\begin{tabular}{c}{$b_1\to a_4\to c_8\to$}\\[-3pt] {$\to a_7\to c_7\to a_1$}\end{tabular}
&6(12)& \\

\hline
$\mtD$ & $ 0 \leqslant\la < \la_1 $ & Yes/No &\begin{tabular}{c}{$b_2\to c_8\to a_7\to$}\\[-3pt] {$\to c_7\to a_1$}\end{tabular}
&6(13)& $I$\,\cite[Table~3]{BRF}\\

\hline
$\mtE_1$ & $ 0 \leqslant\la < 1 $ & Yes/Yes &\begin{tabular}{c}{$b_1\to c_6\to a_5\to$}\\[-3pt] {$\to c_4\to a_2\to a_1$}\end{tabular}&5(8)& \begin{tabular}{c}$D$\,\cite[Table~3]{BRF},\\ \,\cite[Table~8]{Mor}\end{tabular}\\

\hline
$\mtE_2$ & $ 0 \leqslant\la < \la_* $ & Yes/No &\begin{tabular}{c}{$b_1\to c_6\to a_5\to$}\\[-3pt] {$\to a_3\to a_1$}\end{tabular}
&4(7)& $E$\,\cite[Table~3]{BRF}\\

\hline
$\mtE_3$ & $ 0 \leqslant\la < \la_* $ & Yes/No &\begin{tabular}{c}{$b_1\to c_6\to a_8\to$}\\[-3pt] {$\to c_9\to a_3\to a_1$}\end{tabular}
&6(9)& $H$\,\cite[Table~3]{BRF}\\

\hline
$\mtE_4$ & $ 0 <\la < +\infty $ & No/No &\begin{tabular}{c}{$b_1\to c_2\to a_9\to$}\\[-3pt] {$\to a_3\to a_1$}\end{tabular}
&8(21)& \\

\hline
$\mtE_5$ & $ 0 <\la < +\infty $ & No/Yes &\begin{tabular}{c}{$b_1\to c_2\to a_9\to$}\\[-3pt] {$\to c_4\to a_2\to a_1$}\end{tabular}&9(22)& $E$\,\cite[Table~8]{Mor}\\

\hline
$\mtE_6$ & $ \la_* <\la < +\infty $ & No/Yes &\begin{tabular}{c}{$b_1\to c_2\to a_{11}\to$}\\[-3pt] {$\to c_3\to a_2\to a_1$}\end{tabular}&6(11)& $G$\,\cite[Table~8]{Mor}\\

\hline
$\mtF$ & $ \la_* <\la < \la_2$ & No/No &\begin{tabular}{c}{$b_1\to a_6\to c_5\to$}\\[-3pt] {$\to a_{10}\to c_1\to a_1$}\end{tabular}&6(10)& \\

\hline
$\mtG$ & $ \la^* <\la < \sqrt{2}$ & No/Yes &\begin{tabular}{c}{$b_1\to a_6\to c_5\to$}\\[-3pt] {$\to a_{10}\to a_{12}$}\end{tabular}&6(14)& $H$\,\cite[Table~8]{Mor}\\

\hline
$\mtH_1$ & $ \la^* <\la < \sqrt{2}$ & No/No &\begin{tabular}{c}{$b_1\to a_6\to a_2\to$}\\[-3pt] {$\to c_2\to a_{12}$}\end{tabular}&7(19)& \\

\hline
$\mtH_2$ & $ \la_4 <\la < +\infty$ & No/Yes &
\begin{tabular}{c}{$b_1\to
c_2\to a_{12}$}\end{tabular}&2(5)& $J$\,\cite[Table~8]{Mor}\\

\hline
$\mtH_3$ & $ \la_5 <\la < +\infty$ & No/Yes &\begin{tabular}{c}{$b_1\to c_2\to a_{11}\to$}\\[-3pt] {$\to a_{10}\to a_{12}$}\end{tabular}&6(14)& $I$\,\cite[Table~8]{Mor}\\

\hline
\end{tabular}

\normalsize
}
\end{table}

Further, let us distinguish the Fomenko graphs by the orientation along the direction of increasing the integral $K$ and by the stable number of critical circles on the levels of $K$, i.e., by the number that does not change when the point $(\ell,h)$ moves inside its chamber. Then we obtain 22 graphs shown in Fig.~\ref{fig_fomgr},{\it b}. Horizontal dashed segments join the stable pairs of the atoms positioned on the same level.

Note the feature that is not usually mentioned when the identity of the Fomenko graphs is discussed.
There exist {\it typical} points of the plane $O\ell h$ for which some of the critical levels of $K$ are nevertheless unstable, i.e., these levels contain two atoms, which, after small perturbation of $(\ell,h)$ inside the chamber, move to different levels. In this case, not as in the case of splitting of the complex atom, the molecules remain identical in the sense of \cite{BolFom}.
In the problem in question the following levels are unstable $\ell=0,$ $k=1+(h-\la^2/2)^2,$ $h > \la^2/2$. These levels are present in any Fomenko graph of the type $W_{0,h}$ with ${h > \la^2/2}$ (the bounding value $h = \la^2/2$ defines an iso-energetic manifold with a degenerate point and is not considered here). Such graphs appear in the chambers $\mtA_2$, $\mtA_3$, $\mtA_{13}$, $\mtE_1$, $\mtE_5$, $\mtE_6$, $\mtG$, $\mtH_3$ and differ from the neighboring ones with the following property: the upper level of $K$ contains two atoms $A_-$ in the first six cases and four atoms $A_-$ in the last two ones. Under a small perturbation of $\ell$ from the zero level without leaving the chamber the upper level splits to two levels and the atoms belonging to it go to different heights with respect to $k$ (in the last two cases they go by pairs, see the graph 14 in Fig.~\ref{fig_fomgr},{\it b}).

Finally, the complete description of the rough phase topology leads to nine groups of molecules identical in the sense of the definition given in \cite{BolFom}.
Considering the graphs different with respect to the $k$-axis orientation and the number of critical circles on the $k$-levels, these groups contain 22 stable and 6 unstable Fomenko graphs.

Obviously, the identical Fomenko graphs exist even for different topologies of iso-energetic manifolds. To distinguish such graphs, one must use precise classification \cite{FomZ,BolFom}, i.e., to put the marks on the edges of the molecules found. For the majority of the chambers in the enhanced $(\ell,h,\lambda)$-space, the last column in Table~\ref{table5} shows the reference to marked molecules from the same chambers found while investigating the partial cases in the works \cite{BRF,Mor}. When there are no analogues the marks for some edges can be found by deforming the molecule to that investigated earlier and containing the correspondent edge. To obtain the final precise classification one needs to find the marks for those edges which does not have known analogues.

\vspace{3ex}

\small

\makeatletter \@addtoreset{equation}{section}
\@addtoreset{footnote}{section}
\renewcommand{\section}{\@startsection{section}{1}{0pt}{1.3ex
plus 1ex minus 1ex}{1.3ex plus .1ex}{}}

{ 

\renewcommand{\refname}{{\bf References}}

}

\noindent\textbf{Contacts}\\[10pt]
\textit{Mikhail P. Kharlamov}, Doctor of Physics and Mathematics, Professor, Russian Presidential Academy of National Economy and Public Administration, Volgograd Branch,
ul. Gagarina, 8, Volgograd, 400131, Russia\\ [10pt]
E-mail: mharlamov@vags.ru\\ [10pt]
\textit{Pavel E. Ryabov}, Candidate of Physics and Mathematics, Associate Professor, Financial University under the Government of the Russian Federation, Leningradskii prosp., 49, Moscow, 125468, Russia \\[10pt]
E-mail: orelryabov@mail.ru

\end{document}